\documentclass[aps,pre,twocolumn,superscriptaddress,longbibliography]{revtex4-2}
\usepackage{amsmath}
\usepackage{amsfonts}
\usepackage{amssymb}
\usepackage{lmodern,dsfont}
\usepackage{graphicx}
\usepackage[usenames,dvipsnames]{xcolor}
\usepackage{bm}
\usepackage[english=american]{csquotes}
\usepackage{xr}

\newcommand{\Av}[1]{\left\langle #1 \right\rangle}
\newcommand{\av}[1]{\langle #1 \rangle}
\newcommand{\n}{\nonumber}
\newcommand{\nn}{\nonumber \\}

\newcommand{\grad}{\bm{\nabla}}

\renewcommand{\eqref}[1]{Eq.~(\ref{#1})}

\begin{document}
\author{Andreas Dechant}
\affiliation{Department of Physics \#1, Graduate School of Science, Kyoto University, Kyoto 606-8502, Japan}
\author{Shin-ichi Sasa}
\affiliation{Department of Physics \#1, Graduate School of Science, Kyoto University, Kyoto 606-8502, Japan}
\author{Sosuke Ito}
\affiliation{Universal Biology Institute, The University of Tokyo, Tokyo 113-0033, Japan}
\affiliation{JST, PRESTO, Saitama 332-0012, Japan}

\title{Geometric decomposition of entropy production into excess, housekeeping and coupling parts}
\date{\today}

\begin{abstract}
For a generic overdamped Langevin dynamics driven out of equilibrium by both time-dependent and nonconservative forces, the entropy production rate can be decomposed into two positive terms, termed excess and housekeeping entropy.
However, this decomposition is not unique: There are two distinct decompositions, one due to Hatano and Sasa, the other one due to Maes and Neto{\v{c}}n{\`y}.
Here, we establish the connection between these two decompositions and provide a simple, geometric interpretation.
We show that this leads to a decomposition of the entropy production rate into three positive terms, which we call excess, housekeeping and coupling part, respectively.
The coupling part characterizes the interplay between the time-dependent and nonconservative forces.
We also derive thermodynamic uncertainty relations for the excess and housekeeping entropy in both the Hatano-Sasa and Maes-Neto{\v{c}}n{\`y}-decomposition and show that all quantities obey integral fluctuation theorems.
We illustrate the decomposition into three terms using a solvable example of a dragged particle in a nonconservative force field.
\end{abstract}

\maketitle

\section{Introduction} \label{sec-introduction}
Entropy production quantifies the degree of time-reversal symmetry breaking, which is the most fundamental feature of out-of-equilibrium systems.
While equilibrium systems are tightly constrained by the requirement of detailed balance, there are many qualitatively different ways in which a system can be driven out of equilibrium.
Some important examples are varying the parameters of the system according to a time-dependent protocol, inducing currents via an external bias, or the relaxation of a system from an initial nonequilibrium configuration.
In every case, we observe a positive rate of entropy production as long as the system remains out of equilibrium.

In a generic situation, any combination of these archetypal processes can occur: A system may be subject to both external bias and time-dependent driving at the same time.
This naturally raises the question of whether we can separate their effects on the entropy production rate.
More specifically, the entropy production rate can be decomposed into positive contributions which can be associated with different parts of the system \cite{sagawa2012fluctuation, ito2013information,hartich2014stochastic, horowitz2014thermodynamics,  Shi15,Shi15b,Pol17,Bis17} or different types of driving \cite{Lan78,Oon98,Hat01,Rue03,Kom08,Ber13,Mae14}.
In the latter case, the concept of excess (or nonadiabatic) and housekeeping (or adiabatic) entropy production has emerged as a useful concept:
The latter describes the effect of a constant bias, which drives the system into a nonequilibrium steady state, while the former is associated with time-dependent changes in the system.

However, there exist two different approaches, due to Hatano and Sasa \cite{Hat01} and to Maes and Neto{\v{c}}n{\`y} \cite{Mae14}, respectively, which both yield a decomposition of the entropy production rate $\sigma = \sigma^\text{ex} + \sigma^\text{hk}$ into positive excess $\sigma^\text{ex}$ and housekeeping $\sigma^\text{hk}$ parts.
In both cases, $\sigma^\text{ex}$ vanishes in the steady state, while $\sigma^\text{hk}$ vanishes in the absence of nonconservative driving forces like an external bias.
In Ref.~\cite{Dec21}, we recently showed that both decompositions can be derived in terms of a single, geometric approach and that the excess part in the Maes-Neto{\v{c}}n{\`y} approach is always larger than that in the Hatano-Sasa approach.

Here, we build upon these results to show that the geometric formalism allows us to find a decomposition into three nonnegative contributions, $\sigma = \sigma^\text{ex} + \sigma^\text{hk} + \sigma^\text{cp}$.
The excess and housekeeping parts retain their interpretation as being due to driving by time-dependent and nonconservative forces, respectively. 
The third contribution, which we refer to as the coupling part, describes the nontrivial interactions between the two types of driving.
It vanishes only if the time-dependent driving and the nonconservative force act on independent degrees of freedom of the system.
The relation between the two types of decomposition and the existence of the coupling contribution are the main results of this article.

The remainder of the article is structured as follows:
In Section \ref{sec-entropy-langevin}, we review the definition an properties of entropy production for overdamped Langevin dynamics.
The Hatano-Sasa and Maes-Neto{\v{c}}n{\`y} decomposition are introduced and their geometric interpretations provided in Sections \ref{sec-hatano-sasa} and \ref{sec-maes-netocny}, respectively.
In Section \ref{sec-coupling}, we derive the decomposition of the entropy production rate into three terms and discuss the properties of the coupling entropy production.
Section \ref{sec-variational} is devoted to variational expressions for the excess, housekeeping and coupling entropy.
Next, we derive extensions of the thermodynamic uncertainty relation to the different types of excess and housekeeping entropy in Section \ref{sec-tur}.
For the Hatano-Sasa decomposition, it is known that the stochastic excess and housekeeping entropy satisfy an integral fluctuation theorem \cite{Hat01,Spe05}.
We show that a similar result also holds for the Maes-Neto{\v{c}}n{\`y} decomposition in Section \ref{sec-fluctuation-theorem}.
In Section \ref{sec-convergence}, we discuss the interpretation of the excess entropy as a Lyapunov function for the convergence towards the instantaneous steady state.
Finally, in Section \ref{sec-example}, we introduce a solvable model to demonstrate the decomposition of the entropy production rate into three terms and discuss the role of the coupling entropy.

\section{Entropy production in overdamped Langevin dynamics} \label{sec-entropy-langevin}
We consider a system of $d$ overdamped Brownian particles with positions $\bm{x}(t) = (x_1(t),\ldots,x_d(t))$.
These particles are in contact with a viscous environment characterized by the mobility $\mu$ and temperature $T$.
In addition, they are subject to a force field $\bm{F}_t(\bm{x})$, which generally depends on the positions as well as explicitly on time.
This force field may include interactions between the particles, as well as conservative and nonconservative external forces.
The dynamics of the particles are described by the Langevin equation
\begin{align}
\dot{\bm{x}}(t) = \mu \bm{F}_t(\bm{x}(t)) + \sqrt{2 \mu T} \bm{\xi}(t) \label{langevin},
\end{align}
where $\bm{\xi}(t)$ is a vector of mutually independent Gaussian white noises. 
We here set the Boltzmann constant to be unity $k_{\rm B}=1$. 
Equivalently, we may describe the system in terms of its time-dependent probability density $p_t(\bm{x})$, which evolves according to the Fokker-Planck equation \cite{Ris86}
\begin{subequations}
\begin{align}
\partial_t p_t(\bm{x}) &= - \grad \cdot \big(\bm{\nu}_t(\bm{x}) p_t(\bm{x}) \big) \label{continuity} \\
\bm{\nu}_t(\bm{x}) &= \mu \big( \bm{F}_t(\bm{x}) - T \grad \ln p_t(\bm{x}) \big) \label{meanvel} ,
\end{align} \label{fokkerplanck}%
\end{subequations}
with given initial state $p_0$.
Here, $\cdot$ denotes the scalar product in $\mathbb{R}^d$.
The quantity $\bm{\nu}_t(\bm{x})$ is called the local mean velocity and describes the local flows in the system.
We remark that all results obtained in the following can also be generalized to the more general case of a time- and position-dependent diffusion matrix.
If the force acting on the particles is time-independent and conservative, that is $\bm{F}(\bm{x}) = - \grad U(\bm{x})$ with a time-independent potential $U(\bm{x})$, then the probability density converges to the Boltzmann-Gibbs equilibrium density for long times,
\begin{align}
p^\text{eq}(\bm{x}) = p^\text{can}(\bm{x}) = \frac{e^{-\frac{U(\bm{x})}{T}}}{\int d\bm{y} \ e^{-\frac{U(\bm{y})}{T}} } \label{boltzmann-gibbs} .
\end{align}
In the equilibrium state, the local mean velocity vanishes, $\bm{\nu}^\text{eq}(\bm{x}) = 0$, which expresses that the system satisfies detailed balance.
Note that here, and in the following, we assume that the potential is sufficiently confining to give rise to a well-defined steady state.
There are two qualitatively different ways of driving the system out of equilibrium:
One possibility is to consider a conservative, yet time-dependent, force $\bm{F}_t(\bm{x}) = - \grad U_t(\bm{x})$.
Here, the time-dependence of $U_t(\bm{x})$ is imposed via some external protocol.
In this case, if we imagine suspending the time evolution of the protocol, the system will relax to the Boltzmann-Gibbs equilibrium \eqref{boltzmann-gibbs} corresponding to the instantaneous potential $U_t(\bm{x})$.
However, due to the finite rate of change of the potential, the system is kept out of equilibrium and the instantaneous local mean velocity $\bm{\nu}_t(\bm{x})$ is nonzero.
The other possibility is to introduce a time-independent, yet nonconservative force, $\bm{F}(\bm{x}) = - \grad U(\bm{x}) + \bm{F}^\text{nc}(\bm{x})$.
This nonconservative force cannot be written as the gradient of a scalar potential function.
In this case, the system will reach a steady state $p^\text{st}(\bm{x})$ in the long-time limit, however, this steady state does not satisfy detailed balance and gives rise to a nonzero steady-state local mean velocity $\bm{\nu}^\text{st}(\bm{x})$.
In both cases, the breaking of detailed balance is characterized by a positive entropy production, which quantifies the asymmetry between forward and backward transitions in the system.
Specifically, the rate of entropy production for the dynamics \eqref{fokkerplanck} is given by \cite{Sek10,Sei12}
\begin{align}
\sigma_t = \frac{1}{\mu T} \int d\bm{x} \ \big\Vert \bm{\nu}_t(\bm{x}) \big\Vert^2 p_t(\bm{x}) \label{entropy} .
\end{align}
This clearly demonstrates that the existence of a nonzero local mean velocity is equivalent to a positive rate of entropy production.
In general, the system may be driven out of equilibrium due to both types of driving, that is, the force may be time-dependent and nonconservative, $\bm{F}_t(\bm{x}) = - \grad U_t(\bm{x}) + \bm{F}_t^\text{nc}(\bm{x})$, where we also allowed the nonconservative force to depend on time.
For later use, we introduce the inner product between two vector fields $\bm{u}(\bm{x})$ and $\bm{v}(\bm{x})$
\begin{align}
\av{\bm{u},\bm{v}}_p = \frac{1}{\mu T} \int d\bm{x} \ \bm{u}(\bm{x}) \cdot \bm{v}(\bm{x}) p_t(\bm{x}) \label{inner-product} .
\end{align}
As a symmetric inner product, this is linear in each argument $\av{c \bm{u}+\bm{w},\bm{v}}_p = c \av{\bm{u},\bm{v}}_p + \av{\bm{w},\bm{v}}_p$ and positive definite $\av{\bm{v},\bm{v}}_p \geq 0$ with $\av{\bm{v},\bm{v}}_p = 0 \Leftrightarrow \bm{v} = 0$.
Note that $\av{\bm{u},\bm{v}}_p$ implies that the average of $\bm{u} \cdot \bm{v}$ is taken with respect to $p_t(\bm{x})$, where we omit the subscript $t$ in the interest of a more compact notation.
In terms of this inner product, the entropy production rate is
\begin{align}
\sigma_t = \av{\bm{\nu}_t,\bm{\nu}_t}_p .
\end{align}

\section{Hatano-Sasa decomposition} \label{sec-hatano-sasa} 
\begin{figure}
\includegraphics[width=.47\textwidth]{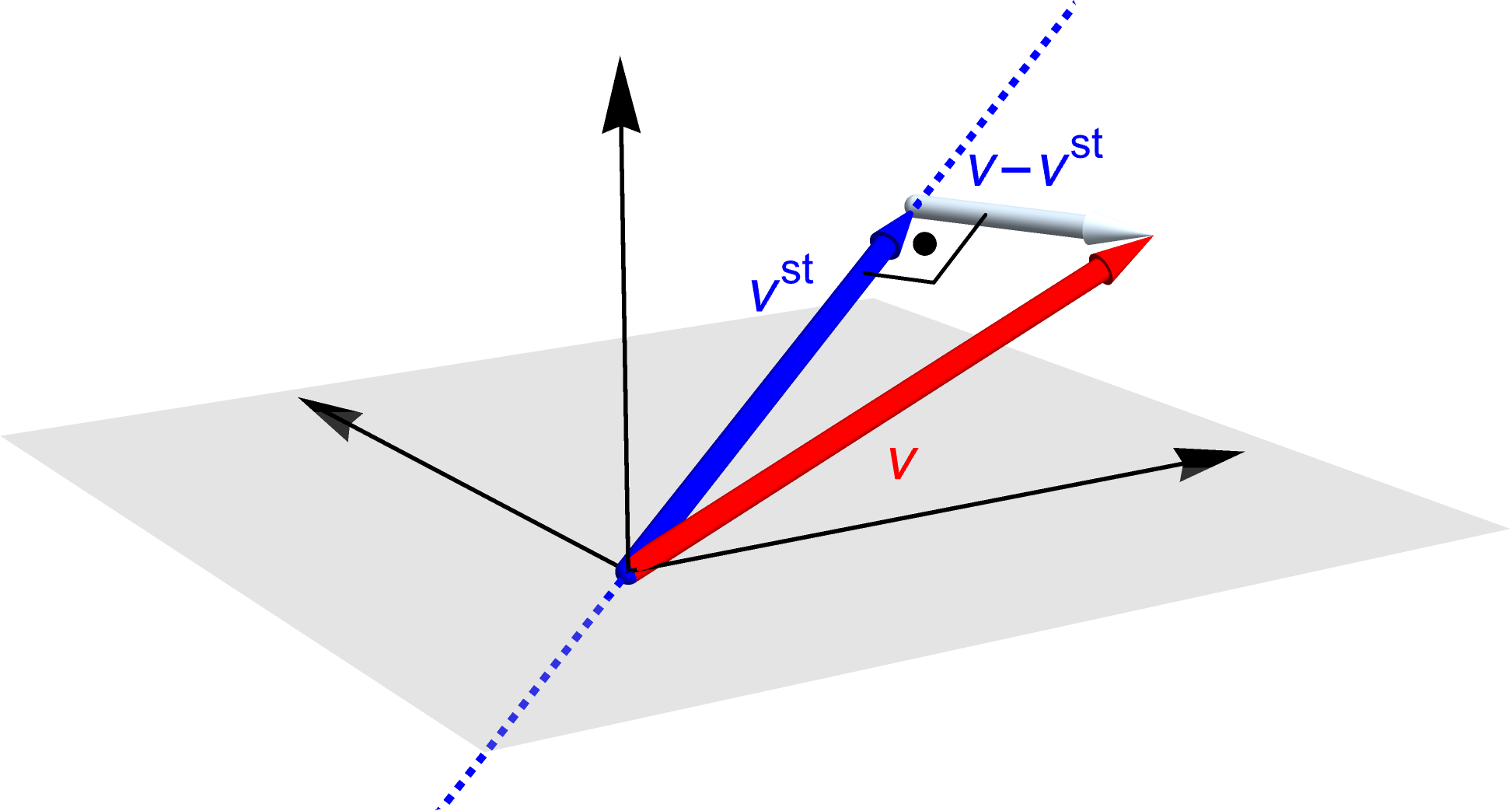}
\caption{Geometric interpretation of the HS decomposition \eqref{HS-decomposition}. The local mean velocity $\bm{\nu}_t(\bm{x})$ can be decomposed into two orthogonal components $\bm{\nu}_t^\text{st}(\bm{x})$ and $\bm{\nu}_t(\bm{x})-\bm{\nu}_t^\text{st}(\bm{x})$, whose length gives the housekeeping and excess entropy production rate, respectively. \label{fig-geometry-hs}}
\end{figure}

The existence of two qualitatively different ways of driving a system out of equilibrium raises the natural question of whether the effect to the two types of driving can be separated.
One way of doing so is provided by the decomposition of the entropy production into excess and housekeeping parts due to Hatano and Sasa \cite{Hat01}.
For the generic case of a time-dependent and nonconservative force, we can still imagine suspending the time evolution of the force at time $t$ and letting the system relax to the steady state corresponding to the instantaneous value of the force.
This steady state obeys the steady-state Fokker-Planck equation
\begin{subequations}
\begin{align}
0 &= - \grad \cdot \big(\bm{\nu}^\text{st}_t(\bm{x}) p^\text{st}_t(\bm{x}) \big) \\
\bm{\nu}^\text{st}_t(\bm{x}) &= \mu \big( \bm{F}_t(\bm{x}) - T \grad \ln p^\text{st}_t(\bm{x}) \big) \label{meanvel-steady} .
\end{align} \label{fokkerplanck-steady}%
\end{subequations}
Note that, since we consider the \emph{instantaneous} steady state of the system, this steady state and the corresponding local mean velocity depend on the time $t$ at which we suspended the time-evolution.
Next, we consider the inner product
\begin{align}
\av{\bm{\nu}_t-\bm{\nu}_t^\text{st},\bm{\nu}_t^\text{st}}_p &= \int d\bm{x} \ \bigg[\grad \ln\frac{p^\text{st}_t(\bm{x})}{p_t(\bm{x})} \bigg] \\
& \qquad \cdot \bm{\nu}_t^\text{st}(\bm{x}) \frac{p_t(\bm{x})}{p_t^\text{st}(\bm{x})} p_t^\text{st}(\bm{x}) , \n
\end{align}
where we inserted a factor $1 = p_t^\text{st}(\bm{x})/p_t^\text{st}(\bm{x})$.
We note that $(\grad \ln f(\bm{x}))/f(\bm{x}) = -\grad (1/f(\bm{x}))$  and thus
\begin{align}
\av{\bm{\nu}_t-\bm{\nu}_t^\text{st},\bm{\nu}_t^\text{st}}_p &= -\int d\bm{x} \ \bigg[ \grad \bigg(\frac{p_t(\bm{x})}{p_t^\text{st}(\bm{x})} \bigg) \bigg] \cdot \bm{\nu}_t^\text{st}(\bm{x}) p_t^\text{st}(\bm{x})  \nn
&= -\mu T \Av{ \grad \bigg(\frac{p_t}{p_t^\text{st}} \bigg), \bm{\nu}_t^\text{st}}_{p^\text{st}} \label{orthogonality-gradient-HS} .
\end{align}
Integrating by parts and using \eqref{fokkerplanck-steady}, we immediately obtain
\begin{align}
\av{\bm{\nu}_t-\bm{\nu}_t^\text{st},\bm{\nu}_t^\text{st}}_p = 0 \label{orthogonality-HS} .
\end{align}
Thus, with respect to the inner product \eqref{inner-product}, the difference between the local mean velocity and its instantaneous steady-state value is orthogonal to the latter.
Then, we can write the entropy production rate as
\begin{align}
\sigma_t = \av{\bm{\nu}_t,\bm{\nu}_t}_p &= \av{\bm{\nu}_t - \bm{\nu}_t^\text{st} + \bm{\nu}_t^\text{st}, \bm{\nu}_t - \bm{\nu}_t^\text{st} + \bm{\nu}_t^\text{st}}_p \nn
&= \av{\bm{\nu}_t - \bm{\nu}_t^\text{st},\bm{\nu}_t - \bm{\nu}_t^\text{st}}_p + \av{\bm{\nu}_t^\text{st}, \bm{\nu}_t^\text{st}}_p \nn
&= \sigma_t^\text{ex,HS} + \sigma_t^\text{hk,HS} \label{HS-decomposition} ,
\end{align}
where we used the linearity of the inner product and \eqref{orthogonality-HS}.
Here the superscript HS denotes the Hatano-Sasa decomposition.
Since each of the two terms is positive, we thus have a decomposition of the entropy production into two positive parts.
The quantity $\sigma_t^\text{ex,HS}$ is the HS excess entropy production rate.
It only vanishes when $p_t = p_t^\text{st}$, that is, when the system is in the steady state at any instant of time.
Thus, generically, we have $\sigma_t^\text{ex,HS} > 0$ whenever the system is driven by a finite-speed protocol.
$\sigma_t^\text{hk,HS}$, by contrast is the HS housekeeping entropy production rate, which only vanishes when the steady-state local mean velocity vanishes, which implies that the instantaneous steady state is in equilibrium.
As a consequence, we have $\sigma_t^\text{hk,HS} > 0$ whenever nonconservative forces are present in the system.
In that sense, \eqref{HS-decomposition} provides a splitting of the entropy production rate into two positive parts which quantify the effects of time-dependent and nonconservative driving, respectively.
Moreover, \eqref{orthogonality-HS} provides a geometric interpretation of this decomposition, as illustrated in Fig.~\ref{fig-geometry-hs}).

\section{Maes-Neto{\v{c}}n{\`y} decomposition} \label{sec-maes-netocny}
\begin{figure}
\includegraphics[width=.47\textwidth]{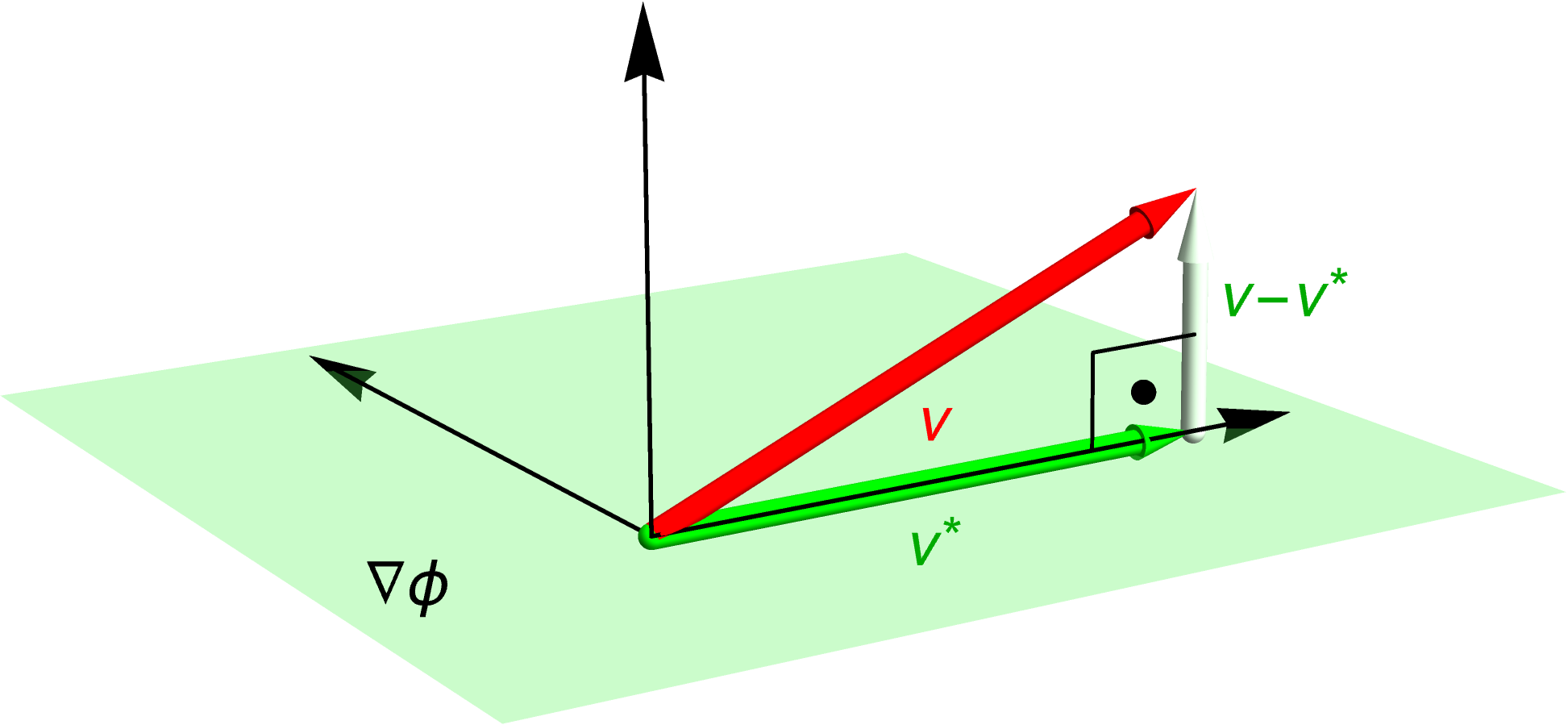}
\caption{Geometric interpretation of the MN decomposition \eqref{MN-decomposition}. The local mean velocity $\bm{\nu}_t(\bm{x})$ can be decomposed into two orthogonal components $\bm{\nu}_t^*(\bm{x})$ and $\bm{\nu}_t(\bm{x})-\bm{\nu}_t^*(\bm{x})$, whose length gives the excess and housekeeping entropy production rate, respectively. Further, the local mean velocity $\bm{\nu}_t^*(\bm{x})$ corresponding to the minimal entropy production rate is obtained as the orthogonal projection of $\bm{\nu}_t(\bm{x})$ into the space of gradient functions (indicated by the shaded plane). \label{fig-geometry-mn}}
\end{figure}

An alternative way of decomposing the entropy production is provided by the minimum entropy production principle introduced by Maes and Neto{\v{c}}n{\`y} \cite{Mae14}. 
The Maes-Neto{\v{c}}n{\`y} decomposition has been recently revisited in Ref.~\cite{Dec21} based on the relationship between stochastic thermodynamics and optimal transport theory~\cite{villani2021topics, aurell2012refined, dechant2019thermodynamic, nakazato2021geometrical}.
Here, the basic idea to search for the force $\bm{F}^*_t(\bm{x})$ and the corresponding local mean velocity $\bm{\nu}_t^*(\bm{x})$ that minimize the entropy production rate \eqref{entropy}, while keeping the time evolution of the probability density $p_t(\bm{x})$ unchanged.
The result is that there exists a unique conservative force $\bm{F}_t^*(\bm{x}) = - \grad U_t^*(\bm{x})$, which gives rise to a given time evolution $p_t(\bm{x})$, that is, for conservative forces, there is a one-to-one correspondence between the force and the probability density.
This conservative force is also the minimizer of the entropy production rate, so that we can write
\begin{align}
\inf_{\bm{F}_t(\bm{x})} \sigma_t = \sigma_t^* = \av{\bm{\nu}_t^*,\bm{\nu}_t^*},
\end{align}
where the optimal local mean velocity is given by
\begin{align}
\bm{\nu}_t^*(\bm{x}) = \mu \big( - \grad U_t^*(\bm{x}) - T \grad \ln p_t(\bm{x}) \big) = - \grad \psi_t^*(\bm{x}) \label{meanvel-optimal},
\end{align}
and is itself the gradient of a scalar function $\psi^*_t(\bm{x})$.
While the fact that $\sigma_t^*$ is obtained by minimizing the entropy production rate already implies that both $\sigma_t^*$ and $\sigma_t - \sigma_t^*$ are positive, this fact can also be shown explicitly.
To see this, we consider the inner product
\begin{align}
&\av{\bm{\nu}_t - \bm{\nu}_t^*, \grad \phi}_p \\
& \qquad = \frac{1}{\mu  T} \int d\bm{x} \ \big[\grad \phi(\bm{x}) \big] \cdot \big( \bm{\nu}_t(\bm{x}) - \bm{\nu}_t^*(\bm{x}) \big) p_t(\bm{x}) \n ,
\end{align}
for some scalar function $\phi(\bm{x})$.
Integrating by parts, we obtain
\begin{align}
&\av{\bm{\nu}_t - \bm{\nu}_t^*, \grad \phi}_p \\
& \qquad = -\frac{1}{\mu  T} \int d\bm{x} \  \phi(\bm{x}) \grad \cdot \big( \bm{\nu}_t(\bm{x}) - \bm{\nu}_t^*(\bm{x}) \big) p_t(\bm{x}) \n .
\end{align}
Since, by definition, both $\bm{\nu}_t(\bm{x})$ and $\bm{\nu}_t^*(\bm{x})$ correspond to the same time evolution, this expression vanishes using \eqref{continuity}.
Thus, we find
\begin{align}
\av{\bm{\nu}_t - \bm{\nu}_t^*, \grad \phi}_p = 0 \label{orthogonality-gradient} ,
\end{align}
which implies that the difference between the local mean velocity and its optimal value is orthogonal to the gradient of any scalar function.
Comparing this to \eqref{orthogonality-gradient-HS}, we see that in the HS decomposition, by contrast, $\bm{\nu}_t^\text{st}(\bm{x})$ is orthogonal only to specific gradient fields; we will come back to this point in Section \ref{sec-variational}.
Since, from \eqref{meanvel-optimal}, we know that the optimal local mean velocity is a gradient field, we obtain
\begin{align}
\av{\bm{\nu}_t-\bm{\nu}_t^*,\bm{\nu}_t^*}_p = 0 \label{orthogonality-MN} ,
\end{align}
in analogy to \eqref{orthogonality-HS}.
This allows us to write the entropy production rate as
\begin{align}
\sigma_t &= \av{\bm{\nu}_t^*,\bm{\nu}_t^*}_p + \av{\bm{\nu}_t-\bm{\nu}_t^*,\bm{\nu}_t-\bm{\nu}_t^*}_p \nn
&= \sigma_t^\text{ex,MN} + \sigma_t^\text{hk,MN} \label{MN-decomposition} .
\end{align}
Here the superscript MN denotes the Maes-Neto{\v{c}}n{\`y} decomposition.
Just like in the HS decomposition, we can decompose the entropy production rate into positive MN excess and housekeeping parts.
The excess entropy production rate $\sigma_t^\text{ex,MN}$ vanishes only if the local mean velocity corresponding to the minimal entropy production rate vanishes, which implies that the system is in a steady state.
Thus, we have $\sigma_t^\text{ex,MN} > 0$ whenever the state of the system depends on time.
Conversely, the housekeeping entropy production rate $\sigma_t^\text{hk,MN}$ vanishes only if the force acting on the system is conservative and thus we have $\sigma_t^\text{hk,MN} > 0$ in the presence of nonconservative forces.
Likewise, the MN decomposition can be given a geometric interpretation using \eqref{orthogonality-MN}, see Fig.~\ref{fig-geometry-mn}.
We remark that one advantage of \eqref{MN-decomposition} is that it does not require the existence of an instantaneous steady state and thus can also be applied, for example, to free diffusion in a nonconservative force field.
We stress that the optimal potential $U_t^*(\bm{x})$ depends on both the force $\bm{F}_t(\bm{x})$ and the probability density $p_t(\bm{x})$:
Even in the same force field, starting from different initial states will generally result in different potential forces.
Intuitively, we can mimic the effects of the force $\bm{F}_t(\bm{x})$ on the present state $p_t(\bm{x})$ by using the conservative force $-\grad U_t^*(\bm{x})$, however, since the two forces are different, this equivalence only holds for the specific state $p_t(\bm{x})$.

\section{Excess, housekeeping and coupling entropy} \label{sec-coupling}
\begin{figure}
\includegraphics[width=.47\textwidth]{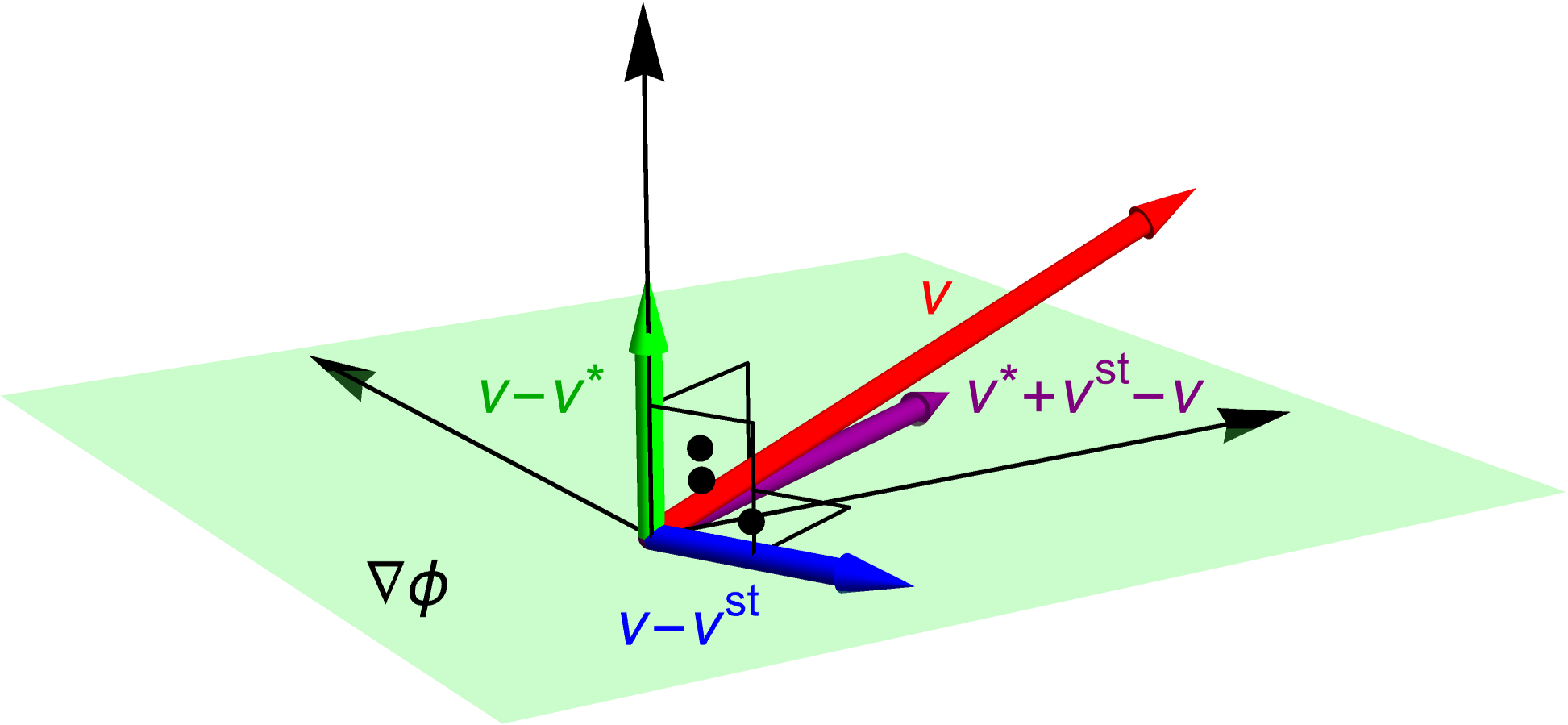}
\caption{Geometric interpretation of the decomposition \eqref{decomposition}. The local mean velocity $\bm{\nu}_t(\bm{x})$ can be decomposed into three orthogonal components $\bm{\nu}_t(\bm{x})-\bm{\nu}_t^\text{st}(\bm{x})$, $\bm{\nu}_t(\bm{x})-\bm{\nu}_t^*(\bm{x})$ and $\bm{\nu}_t^*(\bm{x})+\bm{\nu}^\text{st}_t(\bm{x})-\bm{\nu}_t(\bm{x})$, whose length gives the excess, housekeeping and coupling entropy production rate, respectively. 
Note that both $\bm{\nu}_t(\bm{x}) - \bm{\nu}_t^\text{st}(\bm{x})$ and $\bm{\nu}_t^*(\bm{x})+\bm{\nu}^\text{st}_t(\bm{x})-\bm{\nu}_t(\bm{x})$ are gradient fields. \label{fig-geometry-coupling}}
\end{figure}

Comparing \eqref{HS-decomposition} and \eqref{MN-decomposition}, we note that the two decompositions are generally only equivalent if the system is either in a steady state ($\sigma^\text{st} = \sigma^\text{hk,HS} = \sigma^\text{hk,MN}$) or driven by a time-dependent conservative force ($\sigma_t = \sigma_t^\text{ex,HS} = \sigma_t^\text{ex,MN}$); in both cases the decomposition becomes trivial since one of the two terms vanishes.
However, in the general case of both time-dependent and nonconservative forces, the two decompositions are different, which naturally leads to the question of how they are related.
In order to establish a relation between \eqref{HS-decomposition} and \eqref{MN-decomposition}, we note that
\begin{align}
\bm{\nu}_t(\bm{x})-\bm{\nu}_t^\text{st}(\bm{x}) = \mu T \grad \ln \frac{p_t^\text{st}(\bm{x})}{p_t(\bm{x})},
\end{align}
such that the left-hand side can be written as the gradient of a scalar function.
Using \eqref{orthogonality-gradient}, we immediately find
\begin{align}
\av{\bm{\nu}_t - \bm{\nu}_t^*,\bm{\nu}_t - \bm{\nu}_t^\text{st}}_p = 0 \label{orthogonality-HS-MN} ,
\end{align}
which imposes an orthogonality relation between the two decompositions.
Next, we write the MN excess entropy production rate as
\begin{align}
&\sigma_t^\text{ex,MN} = \av{\bm{\nu}_t^*,\bm{\nu}_t^*}_p \\
&= \av{\bm{\nu}_t^* + \bm{\nu}_t^\text{st} - \bm{\nu}_t - \bm{\nu}_t^\text{st} + \bm{\nu}_t,\bm{\nu}_t^* + \bm{\nu}_t^\text{st} - \bm{\nu}_t - \bm{\nu}_t^\text{st} + \bm{\nu}_t}_p \nn
&= \av{\bm{\nu}_t - \bm{\nu}_t^\text{st},\bm{\nu}_t - \bm{\nu}_t^\text{st}}_p + \av{\bm{\nu}_t^* + \bm{\nu}_t^\text{st} - \bm{\nu}_t,\bm{\nu}_t^* + \bm{\nu}_t^\text{st} - \bm{\nu}_t}_p \nn
&\qquad - 2 \av{\bm{\nu}_t - \bm{\nu}_t^\text{st},\bm{\nu}_t^* + \bm{\nu}_t^\text{st} - \bm{\nu}_t}_p \n .
\end{align}
The first term can be identified as the HS excess entropy production rate and the second term is positive and can be written as
\begin{align}
&\av{\bm{\nu}_t^* + \bm{\nu}_t^\text{st} - \bm{\nu}_t,\bm{\nu}_t^* + \bm{\nu}_t^\text{st} - \bm{\nu}_t}_p \nn
& \; = \av{\bm{\nu}_t^* - \bm{\nu}_t,\bm{\nu}_t^*}_p + \av{\bm{\nu}_t^* - \bm{\nu}_t,\bm{\nu}_t^\text{st} - \bm{\nu}_t}_p \nn
&\qquad + \av{\bm{\nu}_t^\text{st},\bm{\nu}_t^\text{st} - \bm{\nu}_t}_p + \av{\bm{\nu}_t^\text{st},\bm{\nu}_t^*}_p \nn
& \; = \av{\bm{\nu}_t^\text{st},\bm{\nu}_t^*}_p \label{coupling-calculation},
\end{align}
where, in the second line, the first, second and third term vanish because of \eqref{orthogonality-MN}, \eqref{orthogonality-HS-MN} and \eqref{orthogonality-HS}, respectively.
The third term, on the other hand evaluates to
\begin{align}
&\av{\bm{\nu}_t - \bm{\nu}_t^\text{st},\bm{\nu}_t^* + \bm{\nu}_t^\text{st} - \bm{\nu}_t} \\
&\qquad = \av{\bm{\nu}_t - \bm{\nu}_t^\text{st},\bm{\nu}_t^*-\bm{\nu}_t}_p + \av{\bm{\nu}_t - \bm{\nu}_t^\text{st},\bm{\nu}_t^\text{st}}_p = 0 , \n
\end{align}
where the first term vanishes because of \eqref{orthogonality-HS-MN} and the second term because of \eqref{orthogonality-HS}.
Finally, as our first main result, we find that the MN excess entropy production rate can be further decomposed into two positive parts
\begin{align}
\sigma_t^\text{ex,MN} = \sigma_t^\text{ex,HS} + \av{\bm{\nu}_t^\text{st},\bm{\nu}_t^*}_p .
\end{align}
This provides the sought relation between the MN and HS decompositions: 
The excess entropy production rate in the former is always larger than in the latter, with the additional contribution given by the inner product of the steady-state and minimum-entropy-production local mean velocities.
Using this in \eqref{HS-decomposition}, we further find, as our second main result, that we can decompose the entropy production rate into three positive terms
\begin{align}
\sigma_t &= \sigma_t^\text{ex,HS} + \sigma_t^\text{hk,MN} + \sigma_t^\text{cp} \label{decomposition}.
\end{align}
We interpret the first term as the excess entropy production rate; it is equal to the HS excess entropy production rate; it is nonzero whenever the state of the system is time-dependent.
The second term, which is equal to the MN housekeeping entropy production, is interpreted as the housekeeping entropy production rate; it is nonzero whenever the system is driven by a nonconservative force.
The final term quantifies the joint effect of time-dependent and nonconservative driving, and we interpret it as the coupling entropy production rate.
A geometric interpretation of \eqref{decomposition} is provided in Fig.~\ref{fig-geometry-coupling}.
Similar to the HS and MN decomposition, we write the local mean velocity as a sum of orthogonal terms, however, instead of two, we now have three mutually orthogonal contributions.
We remark that both $\bm{\nu}_t(\bm{x}) - \bm{\nu}_t^\text{st}(\bm{x})$ (corresponding to the excess entropy) and $\bm{\nu}_t^*(\bm{x})+\bm{\nu}^\text{st}_t(\bm{x})-\bm{\nu}_t(\bm{x})$ (corresponding to the coupling entropy) are gradient fields, while $\bm{\nu}_t(\bm{x}) - \bm{\nu}_t^*(\bm{x})$ (corresponding to the housekeeping entropy), as argued above, is orthogonal to the space of gradient fields.
The coupling part has two equivalent expressions via \eqref{coupling-calculation},
\begin{align}
\sigma_t^\text{cp} = \av{\bm{\nu}_t^\text{st},\bm{\nu}_t^*}_p = \av{\bm{\nu}_t^* + \bm{\nu}_t^\text{st} - \bm{\nu}_t,\bm{\nu}_t^* + \bm{\nu}_t^\text{st} - \bm{\nu}_t}_p \label{coupling},
\end{align}
where the first expression emphasizes the interpretation as a coupling between the currents $\bm{\nu}_t^*(\bm{x})$ which govern the time evolution of the system and the steady-state currents $\bm{\nu}_t^\text{st}(\bm{x})$ as a consequence of the nonconservative force, while the second expression explicitly demonstrates the positivity.
The coupling entropy production is nonzero only if both sources of nonequilibrium are present and interact in a nontrivial manner.
Specifically, $\sigma_t^\text{cp} = 0$ implies
\begin{align}
\bm{\nu}_t(\bm{x}) = \bm{\nu}_t^\text{st}(\bm{x}) + \bm{\nu}_t^*(\bm{x}) \label{coupling-condition} .
\end{align}
Obviously, this is satisfied if either $\bm{\nu}_t^\text{st}(\bm{x}) = 0$ (implying conservative driving) or $\bm{\nu}_t^*(\bm{x}) = 0$ (implying a steady state).
In principle, this relation may also be satisfied if both time-dependent and nonconservative driving are present but their effects on the system are independent of each other (see below).
However, in the generic case, $\sigma_t^\text{cp} > 0$ in the presence of both time-dependent and nonconservative driving.
In terms of the coupling part of the entropy production rate, the difference between the decompositions \eqref{HS-decomposition} and \eqref{MN-decomposition} is whether this term is included in the housekeeping part (HS decomposition) or the excess part (MN decomposition).
We remark that, in general, explicitly computing the decomposition \eqref{decomposition} is challenging, since we need to know the solution to the Fokker-Planck equation as well as the instantaneous steady state.
However, for systems with linear forces, whose probability density is Gaussian, we can derive more explicit expressions, as is done in Appendix~\ref{app-gaussian}.

The origin of the three terms in \eqref{decomposition} can be clarified further by noting that they may be rewritten as
\begin{subequations}
\begin{align}
\sigma_t^\text{ex,HS} &= - \int d\bm{x} \ \ln \bigg(\frac{p_t(\bm{x})}{p_t^\text{st}(\bm{x})} \bigg) \partial_t p_t(\bm{x}) \\
\sigma_t^\text{cp} &= - \int d\bm{x} \ \ln \bigg(\frac{p_t^\text{st}(\bm{x})}{p_t^\text{can}(\bm{x})} \bigg) \partial_t p_t(\bm{x}) \\
\sigma_t^\text{hk,MN} &= \int d\bm{x} \ \Big\Vert \bm{F}_t(\bm{x}) - \bm{F}_t^*(\bm{x}) \Big\Vert^2 p_t(\bm{x}) .
\end{align} \label{decomposition-explicit}%
\end{subequations}
Here, $\bm{F}_t^*(\bm{x}) = - \grad U_t^*(\bm{x})$ is the force corresponding to the minimum entropy production dynamics \eqref{meanvel-optimal} and we defined the canonical distribution corresponding to $U_t^*(\bm{x})$ as $p_t^\text{can}(\bm{x}) \propto e^{-\frac{U_t^*(\bm{x})}{T}}$.
Thus, the time evolution of the probability density \eqref{fokkerplanck} gives rise to two co-evolving steady states:
The first one, $p_t^\text{st}(\bm{x})$ is obtained by fixing the value of the force $\bm{F}_t(\bm{x})$ and letting the system relax into the (generally) nonequilibrium steady state.
However, since the force $\bm{F}_t^*(\bm{x})$ leads to the same time evolution, we may also replace $\bm{F}_t(\bm{x})$ with $\bm{F}_t^*(\bm{x})$, fix the instantaneous value of $\bm{F}_t^*(\bm{x})$ and then let the system relax, which leads to the equilibrium state $p_t^\text{can}(\bm{x})$.
The two states are the same if the system is either driven by a conservative force (in this case $p_t^\text{st}(\bm{x})$ is already the unique instantaneous equilibrium state) or already in the steady state (in this case, the potential leading to zero minimum entropy production is just $U^*(\bm{x}) = - T \ln p^\text{st}(\bm{x})$).
Let us now return to \eqref{decomposition-explicit}.
The excess entropy production clearly quantifies the difference between the probability density and its co-evolving instantaneous steady state.
The coupling entropy production, on the other hand, quantifies the difference between the co-evolving steady state and equilibrium state, that is, how much the instantaneous steady states of the dynamics driven by $\bm{F}_t(\bm{x})$ and $\bm{F}_t^*(\bm{x})$ differ.
Finally, the housekeeping entropy production directly measures the magnitude of the nonconservative part of the force $\bm{F}_t(\bm{x}) - \bm{F}_t^*(\bm{x})$.
We note that the housekeeping entropy may also be written as
\begin{align}
\sigma_t^\text{hk,MN} = - \int d\bm{x} \ \ln p_t^\text{can}(\bm{x}) \partial_t p_t(\bm{x}) + \frac{\av{\dot{Q}}_t}{T}, 
\end{align}
where $\dot{Q} = \bm{F}_t(\bm{x}(t)) \circ \dot{\bm{x}}(t)$ is the rate at which heat is dissipated into the environment with $\circ$ the Stratonovich product \cite{Sek10,Sei12}.
Using this form together with \eqref{decomposition-explicit} in \eqref{decomposition}, we immediately recover the decomposition of the entropy production rate into the rate of Gibbs-Shannon (GS) entropy change and the dissipated heat
\begin{align}
\sigma_t &= \sigma_t^\text{GS} + \frac{\av{\dot{Q}}_t}{T} \quad \text{with} \\
\sigma_t^\text{GS} &= - d_t \int d\bm{x} \ \ln p_t(\bm{x}) p_t(\bm{x}) \n .
\end{align}
From the positivity of the individual terms in \eqref{decomposition-explicit}, we have the series of inequalities
\begin{align}
\sigma_t^\text{GS} &\geq - \int d\bm{x} \ \ln p_t^\text{st}(\bm{x}) \partial_t p_t(\bm{x}) \\
&\geq - \int d\bm{x} \ \ln p_t^\text{can}(\bm{x}) \partial_t p_t(\bm{x}) \geq - \frac{\av{\dot{Q}}_t}{T} \n ,
\end{align}
or, in terms of the excess heat dissipation rates
\begin{align}
\sigma_t^\text{GS} &\geq - \frac{\av{\dot{Q}^\text{ex,HS}}_t}{T} \geq - \frac{\av{\dot{Q}^\text{ex,MN}}_t}{T} \geq - \frac{\av{\dot{Q}}_t}{T} \label{clausius},
\end{align}
where we defined,
\begin{align}
    \sigma_t^\text{ex,HS} &= \sigma_t^\text{GS} + \frac{\av{\dot{Q}^\text{ex,HS}}_t}{T} \\
    \sigma_t^\text{ex,MN} &= \sigma_t^\text{GS} + \frac{\av{\dot{Q}^\text{ex,MN}}_t}{T} \n .
\end{align}
The outermost inequality represents the usual Clausius inequality between the entropy change of the system and the heat dissipated into the environment.
In the absence of nonconservative forces, we recover an equality in the quasistatic limit, where the state of the system changes slowly and it is in equilibrium at any given time.
However, in the presence of nonconservative forces, which keep dissipating heat even in the steady state, this inequality becomes meaningless in the quasistatic limit.
By identifying the intermediate quantities in \eqref{clausius} as excess heat dissipation rates, we again recover an equality in the limit of slow driving, as has been discussed in Refs.~\cite{Hat01,Mae14}.
Note that, in the quasistatic limit, where the system is assumed to be in the steady state at every instant, we have $p_t^\text{can}(\bm{x}) = p_t^\text{st}(\bm{x})$, so the HS and MN decompositions agree with one another.
For finite-speed driving, on the other hand, the excess dissipation of the MN decomposition is always larger than the HS one.

Finally, we discuss the case where the coupling part in \eqref{decomposition} vanishes, while the excess and housekeeping part remain finite.
We recall that the MN decomposition allows us to view an arbitrary dynamics as the conservative dynamics driven by the potential $U_t^*(\bm{x})$ with an additional nonconservative force $\widetilde{\bm{F}}_t^\text{nc}(\bm{x}) = (\bm{\nu}_t(\bm{x}) - \bm{\nu}_t^*(\bm{x}))/\mu$, which satisfies
\begin{align}
\grad \cdot \big( \widetilde{\bm{F}}_t^\text{nc}(\bm{x}) p_t(\bm{x}) \big) = 0 \label{nc-condition}.
\end{align}
In terms of $U_t^*(\bm{x})$ and $\widetilde{\bm{F}}_t^\text{nc}(\bm{x})$, the instantaneous steady state is determined by
\begin{align}
\grad \cdot \Big( \big( \grad U^*_t(\bm{x}) - \widetilde{\bm{F}}_t^\text{nc}(\bm{x}) + T \grad \ln p_t^\text{st}(\bm{x}) \big) p_t^\text{st}(\bm{x}) \Big) = 0 .
\end{align}
Since, from \eqref{coupling}, we have
\begin{align}
\sigma_t^\text{cp} = \mu^2 T^2 \Av{ \grad \ln \bigg(\frac{p_t^\text{st}}{p_t^\text{can}} \bigg),\grad \ln \bigg(\frac{p_t^\text{st}}{p_t^\text{can}} \bigg)}_p,
\end{align}
a vanishing coupling part implies $p_t^\text{st}(\bm{x}) = p_t^\text{can}(\bm{x}) \propto \exp[-U_t^*(\bm{x})/T]$, which leads to the condition
\begin{align}
\grad \cdot \big( \widetilde{\bm{F}}_t^\text{nc}(\bm{x}) p^\text{can}_t(\bm{x}) \big) = 0 \label{nc-condition-2} .
\end{align}
Comparing this to \eqref{nc-condition}, we obtain the condition on the nonconservative force
\begin{align}
\widetilde{\bm{F}}_t^\text{nc}(\bm{x}) \cdot \grad \ln \bigg(\frac{p_t(\bm{x})}{p^\text{can}_t(\bm{x})} \bigg) = 0 .
\end{align}
This condition is both necessary and sufficient, since we always have \eqref{nc-condition}, which then implies \eqref{nc-condition-2}, and, since the steady state is unique, $p_t^\text{st}(\bm{x}) = p_t^\text{can}(\bm{x})$.
Written in terms of $\bm{\nu}_t(\bm{x})$ and $\bm{\nu}_t^*(\bm{x})$, we have
\begin{align}
\big(\bm{\nu}_t(\bm{x}) - \bm{\nu}_t^*(\bm{x}) \big) \cdot \bm{\nu}_t^*(\bm{x}) = 0 \label{zero-coupling-orthogonal}.
\end{align}
This implies that, for a vanishing coupling part, the orthogonality condition between the flows $\bm{\nu}_t^*(\bm{x})$ contributing to the time evolution and the nonconservative flows $\bm{\nu}_t(\bm{x}) - \bm{\nu}_t^*(\bm{x})$ has to hold not only on the ensemble averaged level of the inner product \eqref{inner-product}, but also on the more microscopic level of the usual inner product in $\mathbb{R}^d$.
\eqref{zero-coupling-orthogonal} shows that, as argued above, the coupling term only vanishes if the time evolution and the nonconservative flows affect separate degrees of freedom and thus do not impact each other.
A vanishing coupling part is equivalent to $\sigma_t^\text{ex,HS} = \sigma_t^\text{ex,MN}$ and $\sigma_t^\text{hk,HS} = \sigma_t^\text{hk,MN}$ and the HS and MN decompositions are identical if and only if the coupling part vanishes.
From the condition $p_t^\text{st}(\bm{x}) = p_t^\text{can}(\bm{x})$, we see that this can only happen if the corresponding components of the local mean velocity coincide,
\begin{align}
    \bm{\nu}_t^*(\bm{x}) = \bm{\nu}_t(\bm{x}) - \bm{\nu}_t^\text{st}(\bm{x}),
\end{align}
and, thus, \eqref{zero-coupling-orthogonal} also implies the microscopic orthogonality relation
\begin{align}
    \big(\bm{\nu}_t(\bm{x}) - \bm{\nu}_t^\text{st}(\bm{x}) \big) \cdot \bm{\nu}_t^\text{st}(\bm{x}) = 0 \label{zero-coupling-orthogonal-HS} .
\end{align}

\section{Variational representation} \label{sec-variational}
In Ref.~\cite{Dec21}, we argued that a decomposition of the local mean velocity into orthogonal components implies variational expressions for the lengths of the individual components.
In this section, we will first re-derive the results in more detail and then apply them to obtain variational formulas for the individual terms in \eqref{decomposition}.
First, we note that the space of local mean velocities associated with a probability density $p_t(\bm{x})$ is a vector space.
More precisely, the space of all smooth vector fields $\bm{v}(\bm{x}): \mathbb{R}^d \rightarrow \mathbb{R}^d$ which are square integrable with respect to $p_t(\bm{x})$, that is $\av{\bm{v},\bm{v}}_p < \infty$, inherits its vector space structure from $\mathbb{R}^d$.
Any such vector field can be interpreted as the instantaneous local mean velocity $\bm{\nu}_t(\bm{x}) = \bm{v}(\bm{x})$ corresponding to some force $\bm{F}_t(\bm{x}) = \bm{v}(\bm{x}) + T \grad \ln p_t(\bm{x})$ via \eqref{fokkerplanck}.
We impose the additional condition that the corresponding steady state exists, that is, we can find a normalized probability density $p_t^\text{st}(\bm{x})$ with $\grad \cdot [(\bm{v}(\bm{x}) - T \grad \ln(p_t^\text{st}(\bm{x})/p_t(\bm{x}))) p_t^\text{st}(\bm{x})] = 0$.
This additional condition keeps the vector space structure intact, and we call the resulting vector space $V$.
Next, we consider a decomposition which separates any such vector field into two orthogonal components, $\bm{v}(\bm{x}) = \bm{v}_1(\bm{x}) + \bm{v}_2(\bm{x})$ with $\av{\bm{v}_1,\bm{v}_2}_p = 0$.
Since this decomposition should be valid for any vector field, this also separates the vector space $V$ into two orthogonal subspaces $V_1$ and $V_2$.
Now, for any $\bm{u}(\bm{x}) \in V_1$ and $\bm{v} \in V$, we have $\av{\bm{u},\bm{v}}_p = \av{\bm{u},\bm{v}_1}_p$.
Then, we use the Cauchy-Schwarz inequality for the inner product,
\begin{align}
\av{\bm{u},\bm{v}}^2 = \av{\bm{u},\bm{v}_1}_p^2 &\leq \av{\bm{u},\bm{u}}_p \av{\bm{v}_1,\bm{v}_1}_p  \nn
\Rightarrow \av{\bm{v}_1,\bm{v}_1}_p &\geq \frac{\av{\bm{u},\bm{v}}_p^2}{\av{\bm{u},\bm{u}}_p} .
\end{align}
Since the choice $\bm{u}(\bm{x}) = c \bm{v}_1(\bm{x})$, with an arbitrary constant $c \neq 0$ results in an equality, we may thus write
\begin{align}
\av{\bm{v}_1,\bm{v}_1}_p = \sup_{\bm{u} \in V_1} \bigg( \frac{\av{\bm{u},\bm{v}}_p^2}{\av{\bm{u},\bm{u}}_p} \bigg) \label{variational-1} .
\end{align}
This is the first general variational expression.
The geometrical interpretation of this expression is that the orthogonal projection $\bm{v}_1(\bm{x})$ of $\bm{v}(\bm{x})$ into the subspace $V_1$ is given by the element $\bm{u}(\bm{x}) \in V_1$ which has the largest overlap with $\bm{v}(\bm{x})$ relative to its length, which is satisfied for any element parallel to $\bm{v}_1(\bm{x})$.
The second variational expression is obtained by noting that for any $\bm{v}(\bm{x}) \in V$ and $\bm{u}(\bm{x}) \in V_2$, we have
\begin{align}
\av{\bm{v} - \bm{u},\bm{v} - \bm{u}}_p &= \av{\bm{v}_1 + \bm{v}_2 - \bm{u},\bm{v}_1 + \bm{v}_2 - \bm{u}}_p \\
&= \av{\bm{v}_1,\bm{v}_1}_p + \av{\bm{v}_2 - \bm{u},\bm{v}_2 - \bm{u}}_p . \n
\end{align}
Since both terms are positive, we have
\begin{align}
\av{\bm{v}_1,\bm{v}_1}_p \leq \av{\bm{v} - \bm{u},\bm{v} - \bm{u}}_p .
\end{align}
Obviously, equality holds for $\bm{u}(\bm{x}) = \bm{v}_2(\bm{x})$, so that we can write
\begin{align}
\av{\bm{v}_1,\bm{v}_1}_p = \inf_{\bm{u} \in V_2} \big( \av{\bm{v} - \bm{u},\bm{v} - \bm{u}}_p \big) \label{variational-2} ,
\end{align}
which is the second general variational expression.
Its geometrical interpretation is that the orthogonal projection $\bm{v}_1(\bm{x})$ of $\bm{v}(\bm{x})$ into the subspace $V_1$ is obtained by minimizing the length of the orthogonal complement.
In principle, we can use the above results to obtain variational representations of both excess and housekeeping entropy production rates, as well as the coupling entropy production rate.
However, since, from the point of view of applications, the expressions for the MN decomposition are most useful, we will focus on the latter in the following; the remaining expressions can be found in Appendix \ref{app-variational}.

In order to apply the general variational expressions \eqref{variational-1} and \eqref{variational-2} to the decompositions \eqref{HS-decomposition}, \eqref{MN-decomposition} and \eqref{decomposition}, we need to identify the relevant subspaces.
For the MN decomposition, this has been done in Ref.~\cite{Dec21}, where it was found that the elements of $V^\text{MN}_1$ and $V^\text{MN}_2$ can be characterized as
\begin{align}
\bm{v}_1(\bm{x}) = \grad \phi(\bm{x}) \quad \text{and} \quad \grad \cdot \big(\bm{v}_2(\bm{x}) p_t(\bm{x}) \big) = 0 \label{space-MN} ,
\end{align}
that is, $V^\text{MN}_1$ is the space of all gradient fields and its orthogonal complement $V^\text{MN}_2$ consists of all vector fields which leave the probability $p_t$ invariant.
We stress that this structure is a consequence of the inner product \eqref{inner-product}, which defines the orthogonality relation; the spaces $V^\text{MN}_1$ and $V^\text{MN}_2$ are generally not orthogonal with respect to the standard inner product on $\mathbb{R}^d$.
Physically, this means that, while there are in general many ways to write the local mean velocity $\bm{\nu}_t(\bm{x})$ as the sum of a gradient and a nongradient term, there exists a unique decomposition such that the nongradient term has no effect on the time evolution of the probability density.
In other words, we may write the local mean velocity as
\begin{align}
\bm{\nu}_t(\bm{x}) &= \mu \big( \bm{F}_t(\bm{x}) - T \grad \ln p_t(\bm{x}) \big) \nn
&= \mu \big( - \grad U_t^*(\bm{x}) + \bar{\bm{F}}_t(\bm{x}) - T \grad \ln p_t(\bm{x}) \big) .
\end{align}
This corresponds to the decomposition of the force $\bm{F}_t$ into a conservative part $\bm{F}_t^*(\bm{x}) = - \grad U_t^*(\bm{x})$ and a nonconservative part $\bar{\bm{F}}_t(\bm{x})$ which satisfies $\grad \cdot ( \bar{\bm{F}}_t(\bm{x}) p_t(\bm{x})) = 0$.
Thus, $\bm{F}_t(\bm{x})$ and $\bm{F}_t^*(\bm{x})$ lead to the same time evolution of $p_t(\bm{x})$, which is unaffected by the nonconservative part.
As was shown in Ref.~\cite{Dec21}, this yields the variational expressions for the excess and housekeeping entropy production rate in the MN decomposition,
\begin{subequations}
\begin{align}
\sigma_t^\text{ex,MN} &= \sup_{\bm{u} \in V^\text{MN}_1} \bigg( \frac{\av{\bm{u}, \bm{\nu}_t}_p^2}{\av{\bm{u},\bm{u}}_p} \bigg) \label{variational-MN-upper}\\
\sigma_t^\text{hk,MN} &= \inf_{\bm{u} \in V^\text{MN}_1} \big( \av{\bm{\nu}_t - \bm{u}, \bm{\nu}_t - \bm{u}}_p \big) \label{variational-MN-lower} .
\end{align} \label{variational-MN}%
\end{subequations}
The advantage of these expressions is that the optimization of the expressions on the right-hand side is carried out over the space $V^\text{MN}_1$ of all gradient fields, which can be done without detailed knowledge about the dynamics.
Therefore, \eqref{variational-MN} is well-suited to determining the excess and housekeeping entropy from trajectory data.
We also have the complementary pair of expressions
\begin{subequations}
\begin{align}
\sigma_t^\text{ex,MN} &= \inf_{\bm{u} \in V^\text{MN}_2} \big( \av{\bm{\nu}_t - \bm{u}, \bm{\nu}_t - \bm{u}}_p \big) \\
\sigma_t^\text{hk,MN} &= \sup_{\bm{u} \in V^\text{MN}_2} \bigg( \frac{\av{\bm{u}, \bm{\nu}_t}_p^2}{\av{\bm{u},\bm{u}}_p} \bigg) .
\end{align} 
\end{subequations}
Since the definition of the space $V^\text{MN}_2$ depends on the probability density $p_t(\bm{x})$ (see \eqref{space-MN}), these expressions are not so useful in practice.

We remark that \eqref{variational-MN} singles out the MN decomposition among all possible decompositions of the local mean velocity into a gradient field and its orthogonal complement,
\begin{align}
\bm{\nu}_t(\bm{x}) = \grad \psi(\bm{x}) + \bm{u}(\bm{x}) \quad \text{with} \quad \av{\grad \psi,\bm{u}}_p = 0 \label{gradient-decomposition}.
\end{align}
Such a decomposition exists for any gradient field $\grad \psi'(\bm{x})$ by choosing $\grad\psi(\bm{x}) = \grad\psi'(\bm{x}) \av{\grad \psi',\bm{\nu}_t}_p/\av{\grad \psi',\grad \psi'}_p$, which is orthogonal to its complement by construction.
Geometrically, this means that we choose some direction in the space $V^\text{MN}_1$ of gradient fields and consider the orthogonal projection of $\bm{\nu}_t(\bm{x})$ onto this direction.
Note that, generally $\bm{u}(\bm{x}) \not\in V_2^\text{MN}$, since the orthogonal complement of a particular gradient field is not necessarily orthogonal any gradient field.
Nevertheless, the orthogonality condition yields a decomposition of the entropy production rate into two positive parts
\begin{align}
\sigma_t = \av{\grad \psi,\grad \psi}_p + \av{\bm{u},\bm{u}}_p .
\end{align}
We may write the second term as
\begin{align}
\av{\bm{u},\bm{u}}_p = \av{\bm{\nu}_t - \grad \psi, \bm{\nu}_t - \grad \psi}_p .
\end{align}
Since $\grad \psi \in V_1^\text{MN}$, this is precisely of the same form as the right-hand side of \eqref{variational-MN-lower}, and we thus have
\begin{align}
\sigma_t^\text{hk,MN} = \av{\bm{u}^*,\bm{u}^*}_p \leq \av{\bm{u},\bm{u}}_p \label{MN-minimum},
\end{align}
where we defined $\bm{u}^*(\bm{x}) = \bm{\nu}_t(\bm{x}) - \bm{\nu}_t^*(\bm{x})$.
This means that, among all decompositions of the local mean velocity into a gradient field and its orthogonal complement \eqref{gradient-decomposition}, the MN decomposition is the one that minimizes the length of the orthogonal complement, or, conversely, maximizes the length of the gradient component.

\section{Thermodynamic uncertainty relations} \label{sec-tur}
The thermodynamic uncertainty relation \cite{Bar15,Gin16,Dec17,Pie17} (TUR) relates the average and fluctuations of a stochastic current to the entropy production in the steady state of a Markov jump or Langevin dynamics.
More recently, this relation has been generalized to time-period \cite{Koy19}, relaxation \cite{Liu20} and arbitrary time-dependent dynamics \cite{Koy20}.
Another recent result is that the HS housekeeping entropy also satisfies a TUR \cite{Chu19b}.
Given the latter result it is natural to ask whether the other terms in the decompositions discussed above also satisfy a similar TUR.

\subsection{Short-time uncertainty relations}
The most straightforward application of the geometric interpretation of the decomposition of the entropy production rate is the derivation of short-time TURs \cite{Dec18,Ots20,Man20,Vu20}.
We define a time-integrated stochastic current as
\begin{align}
J_\tau = \int_0^\tau dt \ \bm{w}_t(\bm{x}(t)) \circ \dot{\bm{x}}(t) \label{current},
\end{align}
where $\bm{w}_t(\bm{x})$ is a weighting function and $\circ$ denotes the Stratonovich product.
The average of such a current is given by
\begin{align}
\av{J_\tau} = \int_0^\tau dt \int d\bm{x} \ \bm{w}_t(\bm{x}) \cdot \bm{\nu}_t(\bm{x}) p_t(\bm{x}) \label{current-average} ,
\end{align}
and its rate of change by
\begin{align}
d_t \av{J_t} = \int d\bm{x} \ \bm{w}_t(\bm{x}) \cdot \bm{\nu}_t(\bm{x}) p_t(\bm{x}) = \mu T \av{\bm{w}_t,\bm{\nu}_t}_p \label{current-rate} .
\end{align}
The short-time TUR can now be readily obtained from the Cauchy-Schwarz inequality for the inner product \cite{Dec18}
\begin{align}
\big(\av{\bm{w}_t,\bm{\nu}_t}_p\big)^2 &\leq \av{\bm{w}_t,\bm{w}_t}_p \av{\bm{\nu}_t,\bm{\nu}_t}_p \nn
\Rightarrow \big(d_t \av{J_t}\big)^2 &\leq \mu T \av{\Vert \bm{w}_t \Vert^2}_t \ \sigma_t \label{short-time-TUR}.
\end{align}
The connection to the TUR, which involves the variance $\text{Var}(J_\tau)$ of the current, is established by noting that the first factor on the right-hand side characterizes the short-time behavior of the latter \cite{Ots20},
\begin{align}
\lim_{\Delta t \rightarrow 0} \frac{\text{Var}(J_{t+\Delta t}-J_t)}{2 \Delta t} = \mu T \av{\Vert \bm{w}_t \Vert^2}_t .
\end{align}
Thus \eqref{short-time-TUR} relates the rate of change of the average current and the short-time growth of its fluctuations to the entropy production rate.
Since the decompositions of the entropy production \eqref{HS-decomposition}, \eqref{MN-decomposition} and \eqref{decomposition} are based on decomposing the local mean velocity into orthogonal components, we can use \eqref{current-rate} to decompose the rate of change of the current accordingly.
For the MN decomposition, the orthogonal spaces are characterized by \eqref{space-MN}.
In particular, if $\bm{w}_t(\bm{x}) = \grad \eta_t(\bm{x})$ is a gradient field, then it is orthogonal to the part of the local mean velocity responsible for the housekeeping entropy,
\begin{align}
\av{\grad \eta_t,\bm{\nu}_t}_p = \av{\grad \eta_t, \bm{\nu}_t^*}_p .
\end{align}
Once again applying the Cauchy-Schwarz inequality, we then obtain, instead of \eqref{short-time-TUR},
\begin{align}
\big(d_t \av{J_t}\big)^2 &\leq \mu T \av{\Vert \grad \eta_t \Vert^2}_t \ \sigma_t^\text{ex,MN} .
\end{align}
Thus, the rate of change of such a current provides a lower bound on the excess part of the entropy production rate; this bound is obviously tighter than \eqref{short-time-TUR}.
This relation can also be obtained from \eqref{variational-MN-lower} by choosing $\grad \eta_t(\bm{x})$ as the gradient field $\bm{u}(\bm{x}) \in V_1^\text{MN}$.
If $\eta(\bm{x})$ is further independent of time, we can identify
\begin{align}
d_t \av{J_t} &= \int d\bm{x} \ \grad \eta(\bm{x}) \cdot \bm{\nu}_t(\bm{x}) p_t(\bm{x}) \\
& = \int d\bm{x} \ \eta(\bm{x}) \partial_t p_t(\bm{x}) = d_t \av{\eta}_t \n ,
\end{align}
where we integrated by parts and used \eqref{fokkerplanck}.
Thus, the change in any scalar observable without explicit time dependence provides a lower bound on the MN excess entropy production rate,
\begin{align}
\big(d_t \av{\eta}_t \big)^2 &\leq \mu T \av{\Vert \grad \eta \Vert^2}_t \ \sigma_t^\text{ex,MN} \label{short-time-TUR-excess-MN}.
\end{align}
Conversely, if the weighting function satisfies $\grad \cdot (\bm{w}_t(\bm{x}) p_t(\bm{x})) = 0$, then we have
\begin{align}
\av{\grad \eta_t,\bm{\nu}_t}_p = \av{\grad \eta_t, \bm{\nu}_t - \bm{\nu}_t^*}_p,
\end{align}
and thus a lower bound on the MN housekeeping entropy,
\begin{align}
\big(d_t \av{J_t}\big)^2 &\leq \mu T \av{\Vert \bm{w}_t \Vert^2}_t \ \sigma_t^\text{hk,MN} \label{short-time-TUR-housekeeping-MN} .
\end{align}
For a general current, we can always decompose its rate of change as
\begin{align}
&d_t \av{J_t} = d_t \av{J_t^\text{ex,MN}} + d_t \av{J_t^\text{hk,MN}} \quad \text{with} \\
&d_t \av{J_t^\text{ex,MN}} = \mu T \av{\bm{w}_t,\bm{\nu}_t^*}_p, \nn
&d_t \av{J_t^\text{hk,MN}} = \mu T \av{\bm{w}_t,\bm{\nu}_t-\bm{\nu}_t^*}_p \n ,
\end{align}
where the individual terms satisfy the inequalities
\begin{subequations}
\begin{align}
\big(d_t \av{J^\text{ex,MN}_t}\big)^2 &\leq \mu T \av{\Vert \bm{w}_t \Vert^2}_t \ \sigma_t^\text{ex,MN}, \\
\big(d_t \av{J^\text{hk,MN}_t}\big)^2 &\leq \mu T \av{\Vert \bm{w}_t \Vert^2}_t \ \sigma_t^\text{hk,MN} .
\end{align} \label{current-decomposition-MN}%
\end{subequations}
Thus, any current can be split into an excess and a housekeeping contribution, which provide lower bounds on the MN excess and housekeeping entropy production rates, respectively.
In a completely analogous manner, we obtain for the HS decomposition,
\begin{align}
&d_t \av{J_t} = d_t \av{J_t^\text{ex,HS}} + d_t \av{J_t^\text{hk,HS}} \quad \text{with} \\
&d_t \av{J_t^\text{ex,HS}} = \mu T \av{\bm{w}_t,\bm{\nu}_t - \bm{\nu}_t^\text{st}}_p, \nn
&d_t \av{J_t^\text{hk,HS}} = \mu T \av{\bm{w}_t,\bm{\nu}_t^\text{st}}_p \n ,
\end{align}
and the lower bounds on the HS excess and housekeeping entropy production rates,
\begin{subequations}
\begin{align}
\big(d_t \av{J^\text{ex,HS}_t}\big)^2 &\leq \mu T \av{\Vert \bm{w}_t \Vert^2}_t \ \sigma_t^\text{ex,HS}, \\
\big(d_t \av{J^\text{hk,HS}_t}\big)^2 &\leq \mu T \av{\Vert \bm{w}_t \Vert^2}_t \ \sigma_t^\text{hk,HS} .
\end{align}
\end{subequations}
Again, if the weighting function is given by $\bm{w}_t = \grad \eta_t(p_t/p_t^\text{st})$, then the housekeeping part vanishes, whereas for $\grad \cdot (\bm{w}_t p_t^\text{st}) = 0$, the excess part vanishes.
The TUR for the HS decomposition is examined in greater detail and extended to the case of Markov jump dynamics in Ref.~\cite{Kam22}.
In principle, we can also split the current into three contributions
\begin{align}
    &d_t \av{J_t} = d_t \av{J_t^\text{ex,HS}} + d_t \av{J_t^\text{hk,MN}} + d_t \av{J_t^\text{cp}}  \\
    & \text{with} \qquad d_t \av{J_t^\text{cp}} = \mu T \av{\bm{w}_t,\bm{\nu}_t^* + \bm{\nu}_t^\text{st} - \bm{\nu}_t}_p \n,
\end{align}
which yields a short-time TUR for the coupling entropy production rate,
\begin{align}
    \big( d_t \Av{J_t^\text{cp}} \big)^2 \leq \mu T \av{\Vert \bm{w}_t \Vert^2}_t \ \sigma_t^\text{cp} .
\end{align}
However, from an operational point of view, that is, for using measured data to estimate the respective entropy production rate, only the short-time TUR for the MN excess entropy \eqref{short-time-TUR-excess-MN} appears immediately applicable, since it only relies on the measurement of the rate of change of scalar observables.
By contrast, measuring the other components of the current requires knowing the respective components of the local mean velocity, which could also be used to calculate the entropy production explicitly.
Thus, the main insight from the remaining TURs is that the current can be decomposed into different contributions corresponding to the orthogonal components of the local mean velocity, whose magnitudes are controlled by the respective contributions to the entropy production rate.

\subsection{Finite-time uncertainty relations for the Maes-Neto{\v{c}}n{\`y} decomposition}
For the MN decomposition, we can extend the above short-time TURs to finite times.
The central observation is that, since $\grad \cdot ((\bm{\nu}_t(\bm{x}) - \bm{\nu}_t^*(\bm{x})) p_t(\bm{x})) = 0$, this term does not contribute to the time evolution of $p_t(\bm{x})$.
As a consequence, we can introduce a modified version of \eqref{langevin}
\begin{align}
\dot{\bm{x}}(t) &= \mu \bm{F}^\theta_t(\bm{x}(t)) + \sqrt{2 \mu T} \bm{\xi}(t) \quad \text{with} \label{langevin-mod} \\
\bm{F}_t^\theta(\bm{x}) &= \bm{F}_t(\bm{x}) + \frac{\theta-1}{\mu} \big( \bm{\nu}_t(\bm{x}) - \bm{\nu}_t^*(\bm{x}) \big) . \n
\end{align}
For $\theta = 1$, this reduces to \eqref{langevin}.
However, for any value of $\theta \in \mathbb{R}$, this dynamics results in the same probability density, that is, the solution of the corresponding Fokker-Planck equation \eqref{fokkerplanck} is given by $p_t^\theta(\bm{x}) = p_t(\bm{x})$ irrespective of $\theta$.
By contrast, the resulting local mean velocity is given by
\begin{align}
\bm{\nu}_t^\theta(\bm{x}) = \theta \bm{\nu}_t(\bm{x}) + (1-\theta) \bm{\nu}_t^*(\bm{x}) .
\end{align}
In particular, for $\theta = 0$, we obtain $\bm{\nu}_t^0(\bm{x}) = \bm{\nu}_t^*(\bm{x})$, that is, the local mean velocity corresponding to the minimum entropy production dynamics driven by a conservative force.
This provides a generalization of the continuous time reversal operation introduced for the steady state in Ref.~\cite{Dec21b}.
In the steady state, we have $\bm{\nu}_t(\bm{x}) = \bm{\nu}_t^\text{st}(\bm{x})$ and $\bm{\nu}_t^*(\bm{x}) = 0$, so \eqref{langevin-mod} interpolates between the original dynamics at $\theta = 1$ and the equilibrium system with the same steady state at $\theta = 0$.
By contrast, if the state of the system depends on time, the interpolation is between the original dynamics at $\theta = 1$ and the dynamics with the same probability density but driven by a conservative force at $\theta = 0$.
The remaining argument proceeds analog to Ref.~\cite{Dec21b}.
We use the fluctuation-response inequality \cite{Dec20}, which relates the change $d\av{X}$ in the average of some quantity $X$ under an infinitesimal perturbation and the variance of $X$ to the Kullback-Leibler divergence between the unperturbed and perturbed probability density.
Since the value of the current \eqref{current} depends on the entire trajectory, we have to consider the ensemble of trajectories $\Gamma$, which is described by the path probability density $\mathbb{P}(\Gamma)$.
As the perturbation, we consider a small change in the parameter $\theta$.
For this case, the fluctuation-response inequality reads
\begin{align}
\frac{\big( \av{J_\tau}_{\theta + d\theta} - \av{J_\tau}_{\theta} \big)^2}{2 \text{Var}(J_\tau)} \leq D_\text{KL}\big(\mathbb{P}^{\theta+d\theta} \Vert \mathbb{P}^{\theta} \big), \label{FRI}
\end{align}
where $D_\text{KL}$ denotes the Kullback-Leibler divergence or relative entropy between the probability densities.
The difference between the averages on the left-hand side is given by
\begin{align}
&\av{J_\tau}_{\theta + d\theta} - \av{J_\tau}_{\theta} \\
&\quad = d\theta \int_0^\tau dt \int d\bm{x} \ \bm{w}_t(\bm{x}) \cdot \big(\bm{\nu}_t(\bm{x}) - \bm{\nu}_t^*(\bm{x}) \big) p_t(\bm{x}) \nn
&\quad = d\theta \av{J^\text{hk,MN}_\tau} \n ,
\end{align}
since the probability density is independent of $\theta$.
Here we identified the housekeeping contribution of the current defined in \eqref{current-decomposition-MN}.
On the other hand, the Kullback-Leibler divergence is given by (see Ref.~\cite{Dec21b} for the calculation)
\begin{align}
D_\text{KL}&\big(\mathbb{P}^{\theta+d\theta} \Vert \mathbb{P}^{\theta} \big) \label{hkfisher} \\
& = \frac{d\theta^2}{4 \mu T} \int_0^\tau dt \int d \bm{x} \ \big\Vert \bm{\nu}_t(\bm{x}) - \bm{\nu}_t^*(\bm{x}) \big\Vert^2 p_t(\bm{x}) \nn
& = \frac{d\theta^2}{4} \int_0^\tau dt \ \av{\bm{\nu}_t-\bm{\nu}_t^*,\bm{\nu}_t-\bm{\nu}_t^*}_p = \frac{d\theta^2}{4} \Delta S_\tau^\text{hk,MN} \n ,
\end{align}
where $\Delta S_\tau^\text{hk,MN} = \int_0^\tau dt \ \sigma_t^\text{hk,MN}$ is the housekeeping entropy production. 
Note that the second-order polynomial of the Kullback-Leibler divergence 
\begin{align}
D_\text{KL}\big(\mathbb{P}^{\theta + d\theta} \big\Vert \mathbb{P}^\theta\big) = \frac{1}{2} g_{\theta\theta} d\theta^2 +O(d\theta^3)
\end{align}
leads to the Fisher metric $g_{\theta \theta} = \int d \Gamma \mathbb{P}^{\theta} (\Gamma) (\partial_{\theta} \ln \mathbb{P}^{\theta} (\Gamma))^2$ in information geometry on the manifold of the path probability~\cite{amari2016information,ito2020unified}. Therefore, the above \eqref{FRI} is essentially the Cram{\'e}r-Rao inequality for the path probability.
The equation (\ref{hkfisher}) implies that the Fisher metric is given by the MN housekeeping entropy production rate 
\begin{align}
   g_{\theta \theta}  = \frac{1}{2}  \Delta S_\tau^\text{hk,MN},
\end{align}
which leads to the finite-time TUR for the MN decomposition. 
Plugging this into \eqref{FRI}, we obtain
\begin{align}
\frac{\av{J^\text{hk,MN}_\tau}^2}{\text{Var}(J_\tau)} \leq \frac{1}{2} \Delta S_\tau^\text{hk,MN} \label{housekeeping-TUR} .
\end{align}
This is precisely the finite-time TUR \cite{Pie16}, with the average current and the entropy production replaced by the respective housekeeping contributions.
While, for time-dependent driving, the form of the TUR has to be modified \cite{Koy19,Koy20}, \eqref{housekeeping-TUR} shows that the original form of the relation is restored by considering the MN housekeeping contributions of the respective quantities.
The downside to \eqref{housekeeping-TUR} is that the housekeeping contribution of the current is generally difficult to evaluate.
We write it as
\begin{align}
\av{J^\text{hk,MN}_\tau} = \av{J_\tau} - \av{J_\tau}^*,
\end{align}
where $\av{J_\tau}^*$ is the average of the current in the minimum entropy production dynamics with the same time evolution as the original dynamics but driven by a conservative force.
Thus, without explicit knowledge of this dynamics, \eqref{housekeeping-TUR} only provides a straightforward way of obtaining a lower bound on the housekeeping entropy if $\av{J_\tau}^*$ vanishes.
As we saw above, the condition for this is that the weighting function satisfies $\grad \cdot (\bm{w}_t(\bm{x}) p_t) = 0$.
Since $\Delta S_\tau^\text{hk,MN} \leq \Delta S_\tau$, such currents also satisfy the conventional TUR; this identifies a class of observables that satisfy the TUR even in the presence of time-dependent driving.
One particular choice for which this is true is $\bm{w}_t(\bm{x}) = \bm{\nu}_t(\bm{x}) - \bm{\nu}_t^*(\bm{x})$.
In this case, the current \eqref{current} can be interpreted as the stochastic housekeeping entropy $\Sigma_\tau^\text{hk,MN}$, for which we obtain
\begin{align}
\frac{\av{\Sigma_\tau^\text{hk,MN}}^2}{\text{Var}(\Sigma_\tau^\text{hk,MN})} \leq \frac{1}{2} \Delta S_\tau^\text{hk,MN} .
\end{align}
Since the average of the stochastic housekeeping entropy is just the housekeeping entropy, we find
\begin{align}
\text{Var}(\Sigma_\tau^\text{hk,MN}) \geq 2 \Delta S_\tau^\text{hk,MN} \label{housekeeping-fano} .
\end{align}
This relation between the fluctuations and the average of the entropy production was previously obtained for the steady state in Ref.~\cite{Pig17} and subsequently extended to the HS housekeeping entropy in Ref.~\cite{Chu19}.
\eqref{housekeeping-fano} shows that the MN housekeeping entropy satisfies the same relation.

We can also obtain a finite-time TUR for the MN excess entropy by generalizing the results of Ref.~\cite{Koy20}.
There, it was shown that the TUR can be extended to time-dependent driving by considering an overall rescaling of time.
We first fix the length of the observation interval $t \in [0,\tau]$ and write the time-dependent forces in the system as
\begin{align}
\bm{F}_t(\bm{x}) = \bar{\bm{F}}_{t/\tau}(\bm{x}) \label{force-reduced} .
\end{align}
We can thus equivalently consider the dynamics of the system in the reduced time $s = t/\tau \in [0,1]$.
For the reduced-time dynamics on the interval $s \in [0,1]$, $\tau$ only enters as a parameter.
This allows us to treat small perturbations in the parameter $\tau$ using \eqref{FRI}, that is, we change $\tau$ while keeping the functional form of the forces fixed, thus also changing the speed of any external protocol.
The second necessary ingredient is that changing $\tau$ in \eqref{fokkerplanck} is the same as rescaling the local mean velocity.
Specifically, the reduced-time version of \eqref{fokkerplanck} reads
\begin{align}
\partial_s p_s^\tau(\bm{x}) &= - \tau \grad \cdot \big( \bm{\nu}_s^\tau(\bm{x}) p_s^\tau(\bm{x}) \big) \quad \text{with} \\
\bm{\nu}_s^\tau(\bm{x}) &= \mu \big(\bar{\bm{F}}_s(\bm{x}) - T \grad \ln p_s^\tau(\bm{x}) \big) \n .
\end{align}
As before, we introduce a modified dynamics by changing the force, $\bar{\bm{F}}_s(\bm{x}) \rightarrow \bar{\bm{F}}_s(\bm{x}) + d\theta \bm{\nu}_s^\tau(\bm{x})/\mu$ with $d\theta \ll 1$, which changes the solution $p_s^\tau(\bm{x}) \rightarrow \tilde{p}_s^\tau(\bm{x})$.
Expanding with respect to $d\theta$, it can be seen that to first order in $d\theta$, we have
\begin{align}
\tilde{p}_s^\tau(\bm{x}) \simeq p_s^{(1+d\theta)\tau}(\bm{x}) + O(d\theta^2) .
\end{align}
Thus, adding a force proportional to the local mean velocity has the same effect as rescaling $\tau$.
For the former perturbation, the Kullback-Leibler divergence in \eqref{FRI} is proportional to the entropy production
\begin{align}
D_\text{KL}\big(\tilde{\mathbb{P}} \Vert \mathbb{P} \big) \simeq \frac{d\theta^2}{4} \Delta S_\tau .
\end{align}
On the other hand, the change in a current \eqref{current} evaluates to
\begin{align}
\widetilde{\av{J}_\tau} - \av{J_\tau} \simeq d\theta \tau \partial_\tau \av{J_\tau} ,
\end{align}
which leads to the generalized TUR \cite{Koy20}
\begin{align}
\frac{\big( \tau \partial_\tau \av{J_\tau} \big)^2}{\text{Var}(J_\tau)} \leq \frac{1}{2} \Delta S_\tau .
\end{align}
Here, the derivative with respect to $\tau$ is understood as the change in the average value of the current when changing the speed of the driving protocol while keeping its functional form \eqref{force-reduced} fixed.
Note that in general, this has two contributions, one from the explicit dependence of \eqref{current} on $\tau$ and one from the implicit dependence of the probability density $p_t$ on the driving speed.
In the steady state, the latter contribution vanishes and thus $\tau \partial_\tau \av{J_\tau} = \av{J_\tau}$, recovering the original finite-time TUR.
The connection to the excess entropy can be made by instead changing the force as $\bar{\bm{F}}_s(\bm{x}) \rightarrow \bar{\bm{F}}_s(\bm{x}) + d\theta \bm{\nu}_s^{\tau,*}(\bm{x})/\mu$, where $\bm{\nu}_s^{\tau,*}(\bm{x})$ is the reduced-time version of the local mean velocity $\bm{\nu}_t^*(\bm{x})$ corresponding to the minimum entropy production dynamics.
In this case, the correspondence between the additional force and the rescaling of time does not hold for arbitrary observables, but only for those whose average depends only on the probability density $p_t$.
From \eqref{current-average}, we see that this is the case whenever the weighting function is a gradient field, $\bm{w}_t(\bm{x}) = \grad \eta_t(\bm{x})$, in which case we have
\begin{align}
\av{J_\tau} &= \int_0^\tau dt \int d\bm{x} \grad \eta_t(\bm{x}) \cdot \bm{\nu}_t(\bm{x}) p_t(\bm{x}) \label{gradient-current} \\
& = \int_0^\tau dt \int d\bm{x} \ \eta_t(\bm{x}) \partial_t p_t(\bm{x}) \n .
\end{align}
These current are precisely the ones whose housekeeping component in \eqref{current-decomposition-MN} vanishes.
The corresponding Kullback-Leibler divergence is proportional to the MN excess entropy, so that we obtain
\begin{align}
\frac{\big( \tau \partial_\tau \av{J_\tau} \big)^2}{\text{Var}(J_\tau)} \leq \frac{1}{2} \Delta S_\tau^\text{ex,MN} \label{excess-TUR}.
\end{align}
This shows that, when the observable is of the form \eqref{gradient-current}, which is in particular true for averages of scalar observables $\eta_t(\bm{x})$, then the TUR derived in Ref.~\cite{Koy20} is actually a lower bound on the MN excess entropy rather than the total entropy production.
Equation~(\ref{excess-TUR}) is useful because it allows us to obtain a lower estimate on the excess entropy production by measuring the change in the average of a state-dependent observable when changing the overall duration of the process.
For example, when a system is in a non-equilibrium steady state and we instantaneously change the parameters of the system, the relaxation towards the new steady state will lead to a finite excess entropy production.
Since, during the relaxation process, the forces in the system remain constant, changing $\tau$ simply amounts to a change in the observation time.
Then, \eqref{excess-TUR} allows us to estimate the excess entropy from the time-dependent relaxation of an arbitrary state-dependent observable.
Note that we may choose time-integrated or instantaneous observables,
\begin{align}
    J_\tau = \int_0^\tau dt \ \eta(\bm{x}(t)) \qquad \text{or} \qquad J_\tau = \eta(\bm{x}(\tau)),
\end{align}
and consider the change in their average with respect to $\tau$; both result in a lower bound on the excess entropy.
On the other hand, if the system is time-periodic with period $\tau$, then changing $\tau$ corresponds to a change in the driving frequency $\omega = 2 \pi/\tau$ \cite{Koy19}.
Again, we may choose time-integrated or instantaneous observables,
\begin{align}
    J_\tau = \int_0^\tau dt \ \eta_t(\bm{x}(t)) \qquad \text{or} \qquad J_\tau = \eta(\bm{x}(r \tau)),
\end{align}
for some fixed $r \in [0,1]$ and where $\eta_{t+\tau}(\bm{x}) = \eta_t(\bm{x})$, and estimate the excess entropy from the change of their averages with respect to $\omega$.

\section{Integral fluctuation theorems} \label{sec-fluctuation-theorem}
\subsection{General formalism}
A remarkable property of the entropy production is that is satisfies an integral fluctuation theorem \cite{Sei05}:
We define the stochastic entropy production as
\begin{align}
\Sigma_\tau =  \int_0^\tau dt \ \bigg(\frac{1}{\mu T} \bm{\nu}_t(\bm{x}(t)) \circ \dot{\bm{x}}(t) - \partial_t \ln p_t(\bm{x}(t)) \bigg) \label{entropy-stochastic} ,
\end{align}
which is a stochastic current of the type \eqref{current}.
Using \eqref{current-average}, we see that the average of $\Sigma_\tau$ is equal to the total entropy production $\Delta S = \int_0^\tau dt \ \sigma_t$ in the time interval $[0,\tau]$.
\eqref{entropy-stochastic} has an equivalent expression in terms of the path probability density,
\begin{align}
\Sigma_\tau = \ln \bigg( \frac{\mathbb{P}(\Gamma)}{\mathbb{P}^\dagger(\Gamma^\dagger)} \bigg) \label{entropy-stochastic-path},
\end{align}
where $\mathbb{P}^\dagger(\Gamma)$ is the time-reversed path probability density.
The trajectory $\Gamma$ describes the time evolution of $\bm{x}(t)$ in \eqref{langevin} in the interval $[0,\tau]$, $\Gamma = (\bm{x}(t))_{t \in [0,\tau]}$.
The time-reversed path probability is obtained from the dynamics \eqref{langevin}, in which the time dependence of the forces is reversed $\bm{F}_t(\bm{x}) \rightarrow \bm{F}_{\tau-t}(\bm{x})$ and which starts from the final state $p_\tau(\bm{x})$.
The trajectory of this dynamics is taken in the time-reversed direction $\Gamma^\dagger = (\bm{x}(\tau-t))_{t \in [0,\tau]}$.
Here, we assume that both path probabilities are evaluated using the mid-point discretization, otherwise $\mathbb{P}^\dagger(\Gamma)$ also involves a change in the discretization scheme \cite{Spi12}.
We note that both $\mathbb{P}(\Gamma)$ and $\mathbb{P}^\dagger(\Gamma^\dagger)$ are normalized path probabilities, $\int d\Gamma \ \mathbb{P}(\Gamma) = \int d\Gamma \ \mathbb{P}^\dagger(\Gamma^\dagger) = 1$.
Using the expression \eqref{entropy-stochastic-path}, it is then straightforward to obtain the integral fluctuation theorem \cite{Sei05,Sei12}
\begin{align}
\Av{e^{-\Sigma_\tau}} = \int d\Gamma \ e^{-\ln \big(\frac{\mathbb{P}^\dagger(\Gamma^\dagger)}{\mathbb{P}(\Gamma)} \big)} \mathbb{P}(\Gamma) = \int d\Gamma \ \mathbb{P}^\dagger(\Gamma^\dagger) = 1 \label{FT} .
\end{align}
This places a strong constraint on the fluctuations of the stochastic entropy production.
More generally, any quantity that can be written as
\begin{align}
\Xi_\tau = \ln \bigg( \frac{\mathbb{P}(\Gamma)}{\tilde{\mathbb{P}}(\Gamma)} \bigg),
\end{align}
with some normalized path probability $\tilde{\mathbb{P}}(\Gamma)$, also satisfies a fluctuation theorem
\begin{align}
\Av{e^{-\Xi_\tau}} = 1 \label{FT-general} .
\end{align}
For the following discussion it is useful to also introduce the modified dynamics
\begin{align}
\dot{\bm{x}}(t) = \mu \bm{F}_t(\bm{x}) + \bm{a}_t(\bm{x}) + \sqrt{2 \mu T} \bm{\xi}(t),
\end{align}
which corresponds to adding a force $\bm{a}_t(\bm{x})/\mu$ to \eqref{langevin}, starting from the same initial state $p_0(\bm{x})$.
We denote the path probability generated by this dynamics as $\mathbb{P}^{a}(\Gamma)$ and its time-reversed version as $\mathbb{P}^{a,\dagger}(\Gamma^\dagger)$, which involves a time reversal of the force, $\bm{F}_{\tau-t}(\bm{x}) + \bm{a}_{\tau-t}(\bm{x})/\mu$, and the trajectory, while starting from the final state of \eqref{langevin}, $p_\tau(\bm{x})$.
For this type of path probability density, we obtain \cite{Ben96} (see Appendix~\ref{app-KL-derivation} for the detailed calculation)
\begin{align}
\Xi^a_\tau = \ln \bigg(\frac{\mathbb{P}(\Gamma)}{\mathbb{P}^a(\Gamma)} \bigg) &= \frac{1}{4 \mu T} \int_0^\tau dt \ \Big( \Vert \bm{a}_t(\bm{x}(t)) \Vert^2 \label{xi-a} \\
&\quad + 2 \bm{a}_t(\bm{x}(t)) \cdot \big( \mu \bm{F}_t(\bm{x}(t)) - \dot{\bm{x}}(t) \big) \Big) \n ,
\end{align}
where the scalar product $\cdot$ is interpreted as an Ito-product.
Using \eqref{langevin}, we see that the average of this quantity is given by
\begin{align}
\av{\Xi^a_\tau} = \frac{1}{4} \int_0^\tau dt \ \av{\bm{a}_t,\bm{a}_t}_p \label{xi-a-average} .
\end{align}
Since the average is equal to the Kullback-Leibler divergence between $\mathbb{P}(\Gamma)$ and $\mathbb{P}^a(\Gamma)$ it is positive.
For the time-reversed path probability density, we find (see Appendix~\ref{app-KL-derivation})
\begin{align}
\Xi^{a,\dagger}_\tau &= \ln \bigg(\frac{\mathbb{P}(\Gamma)}{\mathbb{P}^{a,\dagger}(\Gamma^\dagger)} \bigg) = \Sigma_\tau \label{xi-a-reverse} \\
&\quad + \frac{1}{4 \mu T} \int_0^\tau dt \ \Big( \Vert \bm{a}_t(\bm{x}(t)) \Vert^2 + 4 \bm{a}_t(\bm{x}(t)) \circ \dot{\bm{x}}(t) \nn
&\quad + 2 \bm{a}_t(\bm{x}(t)) \cdot \big( \mu \bm{F}_t(\bm{x}(t)) - \dot{\bm{x}}(t) \big)  \Big) \n .
\end{align}
For $\bm{a}_t(\bm{x}) = 0$, we recover \eqref{entropy-stochastic-path}.
Taking the average of this quantity, we find
\begin{align}
\av{\Xi^{a,\dagger}_\tau} &= \Delta S + \frac{1}{4} \int_0^\tau dt \ \Big(\av{\bm{a}_t,\bm{a}_t}_p + 4 \av{\bm{a}_t,\bm{\nu}_t}_p \Big) \label{xi-a-reverse-average} \\
&= \int_0^\tau dt \ \Av{\bm{\nu}_t + \frac{1}{2} \bm{a}_t,\bm{\nu}_t + \frac{1}{2} \bm{a}_t}_p \n .
\end{align}
Again, this is positive and attains its minimal value of $0$ for $\bm{a}_t(\bm{x}) = - 2\bm{\nu}_t(\bm{x})$.
Note that \eqref{xi-a-average} implies that, for any dimensionless positive quantity $B > 0$, we can find arbitrarily many stochastic observables whose average is $B$, and all of which obey an integral fluctuation theorem.
Specifically, for any $\bm{a}_t(\bm{x})$, we may choose $\tilde{\bm{a}}_t(\bm{x}) = \sqrt{4B/\av{\Xi_\tau^a}} \bm{a}_t(\bm{x})$ and have
\begin{align}
    \Av{ e^{-\Xi_\tau^{\tilde{a}} } } = 1 \qquad \text{and} \qquad \av{\Xi_\tau^{\tilde{a}}} = B .
\end{align}

\subsection{HS decomposition}
We now choose $\bm{a}_t(\bm{x}) = -2 \bm{\nu}^\text{st}_t(\bm{x})$ in \eqref{xi-a}, which is also referred to as the dual dynamics \cite{Che06,Sei12,Sas14}.
Then, using \eqref{xi-a-average}, we immediately have $\av{\Xi^a_\tau} = \Delta S^\text{hk,HS}$.
So, we have a stochastic quantity, whose average is equal to the HS housekeeping entropy and which satisfies the integral fluctuation theorem \eqref{FT-general}.
This quantity can be rewritten as
\begin{align}
\Sigma_\tau^\text{hk,HS} = \frac{1}{\mu T} \int_0^\tau dt \ \bm{\nu}_t^\text{st}(\bm{x}(t)) \circ \dot{\bm{x}}(t) ,
\end{align}
which is a natural definition of the stochastic HS housekeeping entropy in light of \eqref{entropy-stochastic}.
In deriving the above, we used the relation between Ito and Stratonovich product,
\begin{align}
\bm{a}_t(\bm{x}(t)) \circ \dot{\bm{x}}(t)  = \bm{a}_t(\bm{x}(t)) \cdot \dot{\bm{x}}(t) + \mu T \grad \cdot \bm{a}_t(\bm{x}(t)) .
\end{align}
The same choice in \eqref{xi-a-reverse} yields the HS excess entropy as an average using \eqref{xi-a-reverse-average} and satisfies an integral fluctuation theorem .
We can rewrite \eqref{xi-a-reverse} as
\begin{align}
\Sigma_\tau^\text{ex,HS} =  \int_0^\tau dt \ \bigg( \frac{1}{\mu T} &\big(\bm{\nu}_t(\bm{x}(t)) - \bm{\nu}_t^\text{st}(\bm{x}(t))\big) \circ \dot{\bm{x}}(t) \nn
&- \partial_t \ln p_t(\bm{x}(t)) \bigg) ,
\end{align}
which is a natural definition of the stochastic HS excess entropy.
Thus, the HS decomposition of the local mean velocity $\bm{\nu}_t(\bm{x}) = \bm{\nu}_t(\bm{x}) - \bm{\nu}_t^\text{st}(\bm{x}) + \bm{\nu}_t^\text{st}(\bm{x})$ in \eqref{entropy-stochastic} also yields a decomposition of the stochastic entropy production, in which both terms satisfy an integral fluctuation theorem \cite{Hat01,Spe05,Spi12b},
\begin{align}
\Av{e^{-\Sigma_\tau^\text{ex,HS}}} = 1, \qquad \Av{e^{-\Sigma_\tau^\text{hk,HS}}} = 1 .
\end{align}
We remark that the integral fluctuation theorems for the HS excess and housekeeping entropy are established in the literature \cite{Che06,Gar12,Sei12}; we provide them here as a reference facilitate the comparison to the results for the MN excess and housekeeping, as well as the coupling entropy, which we derive below.
We further note that the decomposition of the stochastic entropy production is not unique; we could equally well choose $\bm{a}_t(\bm{x}) = -2 (\bm{\nu}_t(\bm{x}) - \bm{\nu}_t^\text{st}(\bm{x}))$, which yields a pair of distinct stochastic quantities, which differ from the above expressions by a term whose average vanishes,
\begin{align}
&\widetilde{\Sigma}_\tau^\text{hk,HS} - \Sigma_\tau^\text{hk,HS} = \widetilde{\Sigma}_\tau^\text{ex,HS} - \Sigma_\tau^\text{ex,HS} \\
& \; = 2 \int_0^\tau dt \ \Big( \partial_t \ln p_t(\bm{x}(t)) \nn
& \qquad - \frac{1}{\mu T} \big(\bm{\nu}_t(\bm{x}(t)) - \bm{\nu}_t^\text{st}(\bm{x}(t)) \big) \cdot \bm{\nu}_t^\text{st}(\bm{x}(t)) \Big) . \n
\end{align}
Both definitions yield the correct averages and satisfy an integral fluctuation theorem.

\subsection{MN decomposition}
For the MN decomposition, we choose $\bm{a}_t(\bm{x}) =-2( \bm{\nu}_t(\bm{x}) - \bm{\nu}_t^*(\bm{x}))$, which from \eqref{xi-a-average} and \eqref{xi-a-reverse-average} yields stochastic quantities whose average is the MN housekeeping and excess entropy, respectively, and both of which satisfy the integral fluctuation theorem.
Explicitly, these can be written as
\begin{subequations}
\begin{align}
\widetilde{\Sigma}_\tau^\text{hk,MN} &= \frac{1}{\mu T} \int_0^\tau dt \ \Big( \big(\bm{\nu}_t(\bm{x}(t)) - \bm{\nu}_t^*(\bm{x}(t)) \big) \circ \dot{\bm{x}}(t) \nn
&\quad - \bm{\nu}_t^*(\bm{x}(t)) \cdot \big(\bm{\nu}_t(\bm{x}(t)) - \bm{\nu}_t^*(\bm{x}(t)) \big) \Big) , \\
\widetilde{\Sigma}_\tau^\text{ex,MN} &= \int_0^\tau dt \ \bigg( \frac{1}{\mu T}  \Big( \bm{\nu}_t^*(\bm{x}(t)) \circ \dot{\bm{x}}(t) \nn
&\quad - \bm{\nu}_t^*(\bm{x}(t)) \cdot \big(\bm{\nu}_t(\bm{x}(t)) - \bm{\nu}_t^*(\bm{x}(t)) \big) \Big) \\
&\qquad \qquad - \partial_t \ln p_t(\bm{x}(t)) \bigg) \n . 
\end{align} \label{entropy-stochastic-MN}%
\end{subequations}
These expressions differ from the naive expressions obtained by replacing $\bm{\nu}_t(\bm{x})$ in \eqref{entropy-stochastic} with $\bm{\nu}_t(\bm{x})-\bm{\nu}_t^*(\bm{x})$ or $\bm{\nu}_t^*(\bm{x})$ by an additional term, whose average vanishes.
In particular, the stochastic entropy production is decomposed as
\begin{align}
\Sigma_\tau &= \widetilde{\Sigma}_\tau^\text{hk,MN} + \widetilde{\Sigma}_\tau^\text{ex,MN} \\
&\quad + \frac{2}{\mu T} \int_0^\tau dt \ \bm{\nu}_t^*(\bm{x}(t)) \cdot \big(\bm{\nu}_t(\bm{x}(t)) - \bm{\nu}_t^*(\bm{x}(t)) \big) \n .
\end{align}
From \eqref{orthogonality-MN}, it is clear that the additional term is zero on average.
However, the presence of this term is necessary in order for the stochastic excess and housekeeping entropy to satisfy the integral fluctuation theorems,
\begin{align}
\Av{e^{-\widetilde{\Sigma}_\tau^\text{ex,MN}}} = 1, \qquad \Av{e^{-\widetilde{\Sigma}_\tau^\text{hk,MN}}} = 1 .
\end{align}
We remark that, even though \eqref{entropy-stochastic-MN} is different from the definition of the stochastic housekeeping entropy used in \eqref{housekeeping-fano}, the latter relation also holds for \eqref{entropy-stochastic-MN}.
To see this, note that the additional term in \eqref{entropy-stochastic-MN} is not a stochastic current of the type \eqref{current}, and thus its average is independent of the parameter $\theta$ introduced in \eqref{langevin-mod}.
As with the HS decomposition, we may also obtain alternative stochastic excess and housekeeping entropy productions by choosing $\bm{a}_t(\bm{x}) = \bm{\nu}_t^*(\bm{x})$.
In summary, both the HS and MN decomposition yields excess and housekeeping contributions to the stochastic entropy production which satisfy an integral fluctuation theorem.
However, the conditions that the respective component of the stochastic entropy production should yield the correct average and satisfy a fluctuation theorem are not sufficient to uniquely specify the decomposition.

\subsection{Coupling entropy}
Finally, choosing $\bm{a}_t(\bm{x}) = -2( \bm{\nu}_t^*(\bm{x}) + \bm{\nu}_t^\text{st}(\bm{x}) - \bm{\nu}_t(\bm{x}))$, we obtain a stochastic version of the coupling entropy
\begin{align}
    \Sigma_\tau^\text{cp} &=   \frac{1}{\mu T} \int_0^\tau dt \Big( \big( \bm{\nu}_t^* +  \bm{\nu}_t^\text{st} - \bm{\nu}_t \big) \circ \dot{\bm{x}} \\
&\quad - \bm{\nu}_t^* \cdot \big(\bm{\nu}_t - \bm{\nu}_t^* \big) - 2 \big( \bm{\nu}_t^* +  \bm{\nu}_t^\text{st} - \bm{\nu}_t \big) \cdot \big( \bm{\nu}_t - \bm{\nu}_t^\text{st} \big) \Big) , \n
\end{align}
where we omitted the arguments for brevity.
This quantity satisfies the integral fluctuation theorem
\begin{align}
    \Av{e^{-\Sigma_\tau^\text{cp}}} = 1 ,
\end{align}
however, just like the stochastic counterpart of the MN decomposition, this definition differs from the \enquote{naive} definition of a stochastic coupling entropy by a term that averages to zero. 
The stochastic entropy production is written as
\begin{align}
    \Sigma_\tau &= \Sigma_\tau^\text{ex,HS} + \widetilde{\Sigma}_\tau^\text{hk,MN} + \Sigma_\tau^\text{cp} \\
    &\quad + \frac{2}{\mu T} \int_0^\tau dt \ \Big(\big( \bm{\nu}_t - \bm{\nu}_t^* \big) \cdot \big(\bm{\nu}_t^* + \bm{\nu}_t^\text{st} - \bm{\nu}_t \big) \nn
    &\hspace{3cm} + \bm{\nu}_t^\text{st} \cdot \big( \bm{\nu}_t - \bm{\nu}_t^\text{st} \big) \Big) \n ,
\end{align}
which likewise contains an additional term that averages to zero and is necessary to ensure that the three contributions each satisfy an integral fluctuation theorem.

\section{Convergence towards the instantaneous steady state} \label{sec-convergence}
A well-known result for the time evolution of the probability density is that, given two solutions $p_t(\bm{x})$ and $q_t(\bm{x})$ of \eqref{fokkerplanck} the Kullback-Leibler (KL) divergence between them,
\begin{align}
D_\text{KL}(p_t \Vert q_t) = \int d\bm{x} \ \ln \bigg(\frac{p_t(\bm{x})}{q_t(\bm{x})} \bigg) p_t(\bm{x}),
\end{align}
is a monotonically decreasing function of time \cite{Leb57,Sch71,Ris86}.
Specifically,
\begin{align}
d_t D_\text{KL}(p_t \Vert q_t) = - \mu T \int d\bm{x} \ \bigg\Vert \grad \ln \bigg(\frac{p_t(\bm{x})}{q_t(\bm{x})} \bigg) \bigg\Vert^2 p_t(\bm{x}) \leq 0 \label{KL-convergence} .
\end{align}
Intuitively, this ensures that, even when starting from different initial conditions, the solution of \eqref{fokkerplanck} converges to a unique limiting solution in the long-time limit.
In other words, the KL divergence serves as a Lyapunov function for the Fokker-Planck equation, guaranteeing the stability of the solution.
Similar to the derivation of the Hatano-Sasa decomposition, we now fix the force at its instantaneous value $\bm{F}_s(\bm{x})$.
In this case, for sufficiently short time $dt = t-s$, both $p_t(\bm{x})$ and $p_s^\text{st}(\bm{x})$ are instantaneous solutions of the corresponding Fokker-Planck equation.
We then have from \eqref{KL-convergence}
\begin{align}
\Big[d_t D_\text{KL}(p_t \Vert p_s^\text{st})\Big]_{t = s} \leq 0 \label{KL-convergence-HS} . 
\end{align}
The force is generally nonconservative, but, as discussed in the derivation of the Maes-Neto{\v{c}}n{\`y} decomposition, we can find a conservative force $\bm{F}_t^*(\bm{x})$ that leads to the same time evolution.
Fixing the value of $\bm{F}_s^*(\bm{x})$, both $p_t$ ($t = s + dt$) and $p_s^\text{can}(\bm{x})$ are instantaneous solutions, so we also have
\begin{align}
\Big[d_t D_\text{KL}(p_t \Vert p_s^\text{can}) \Big]_{t = s} \leq 0 \label{KL-convergence-MN} .
\end{align}
Using the explicit expression \eqref{KL-convergence}, we find
\begin{subequations}
\begin{align}
\Big[d_t D_\text{KL}(p_t \Vert p_s^\text{st})\Big]_{t = s} &= - \sigma_t^\text{ex,HS}, \\
\Big[d_t D_\text{KL}(p_t \Vert p_s^\text{can}) \Big]_{t = s} &= - \sigma_t^\text{ex,MN}.
\end{align}
\end{subequations}
This means that the excess entropy production rates serve as Lyapunov functions for the dynamics with the force $\bm{F}_t(\bm{x})$ and $\bm{F}_t^*(\bm{x})$, respectively.
Further, the inequality $\sigma_t^\text{ex,HS} \leq \sigma_t^\text{ex,MN}$ implies that the instantaneous rate at which the respective limit distribution is approached is faster for the system driven by the conservative force $\bm{F}_t^*(\bm{x})$.
We remark that this does not contradict the finding that nonconservative forces generally lead to a faster approach towards the steady state \cite{Ohz15}, since the limit distributions for $\bm{F}_t(\bm{x})$ and $\bm{F}_t^*(\bm{x})$ are not the same.
For a relaxation process with time-independent force, we obtain \cite{Ge09}
\begin{align}
d_t D_\text{KL}(p_t \Vert p^\text{st}) = - \sigma_t^\text{ex,HS} ,
\end{align}
and thus the HS excess entropy production rate characterizes the approach to the steady state.
\eqref{KL-convergence-HS} is the generalization of this result to time-dependent forces, which follows by noting that $p_s^\text{st}(\bm{x})$ is the steady state corresponding to the force $\bm{F}_s(\bm{x})$ with fixed $s$.
Note that, even if the force in the original dynamics is time-independent, the force $\bm{F}_t^*(\bm{x})$ does depend on time and so does $p_t^\text{can}(\bm{x})$.

\section{Demonstration} \label{sec-example}
\subsection{Periodic change in the position of a parabolic trap}
\begin{figure}
\includegraphics[width=.47\textwidth]{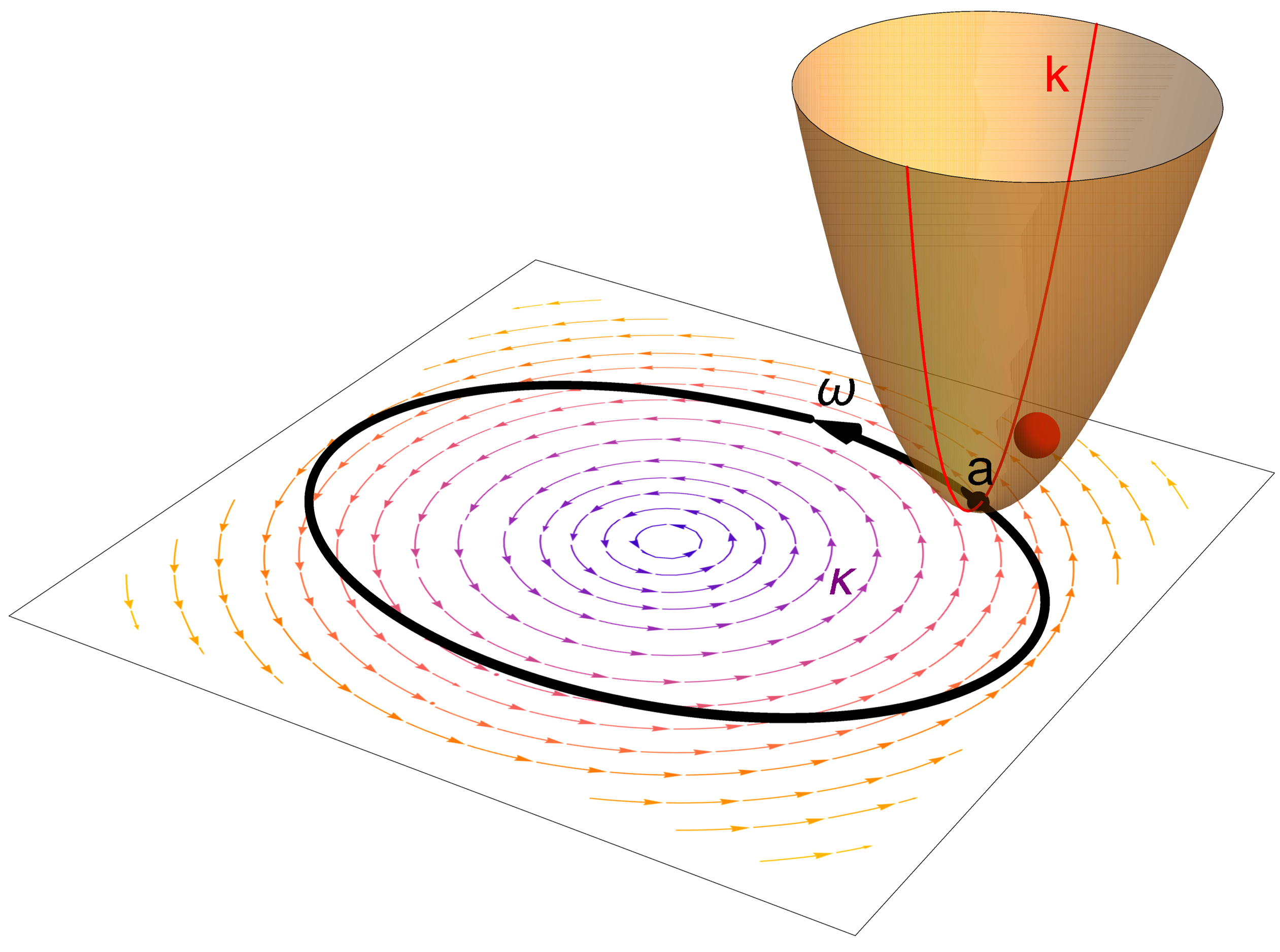}\\
\caption{A graphical illustration of the example system. A particle is trapped in a parabolic potential with spring constant $k$, whose minimum position $\bm{a}_t$ moves around the origin with angular frequency $\omega$. In addition, the particle is driven by a nonconservative force field of strength $\kappa$. \label{fig-example}}
\end{figure}
\begin{figure}
\includegraphics[width=.47\textwidth]{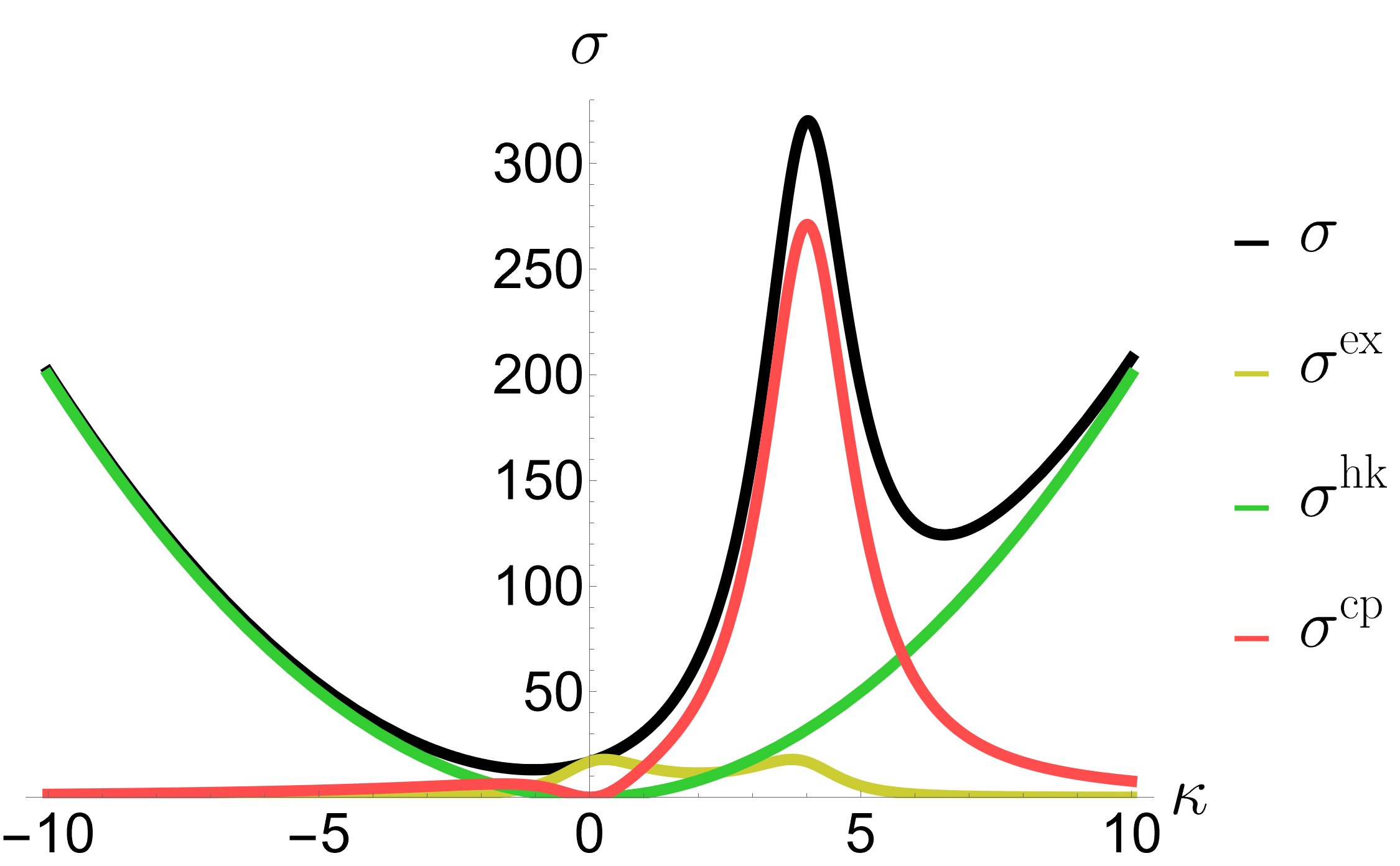}\\
\includegraphics[width=.47\textwidth]{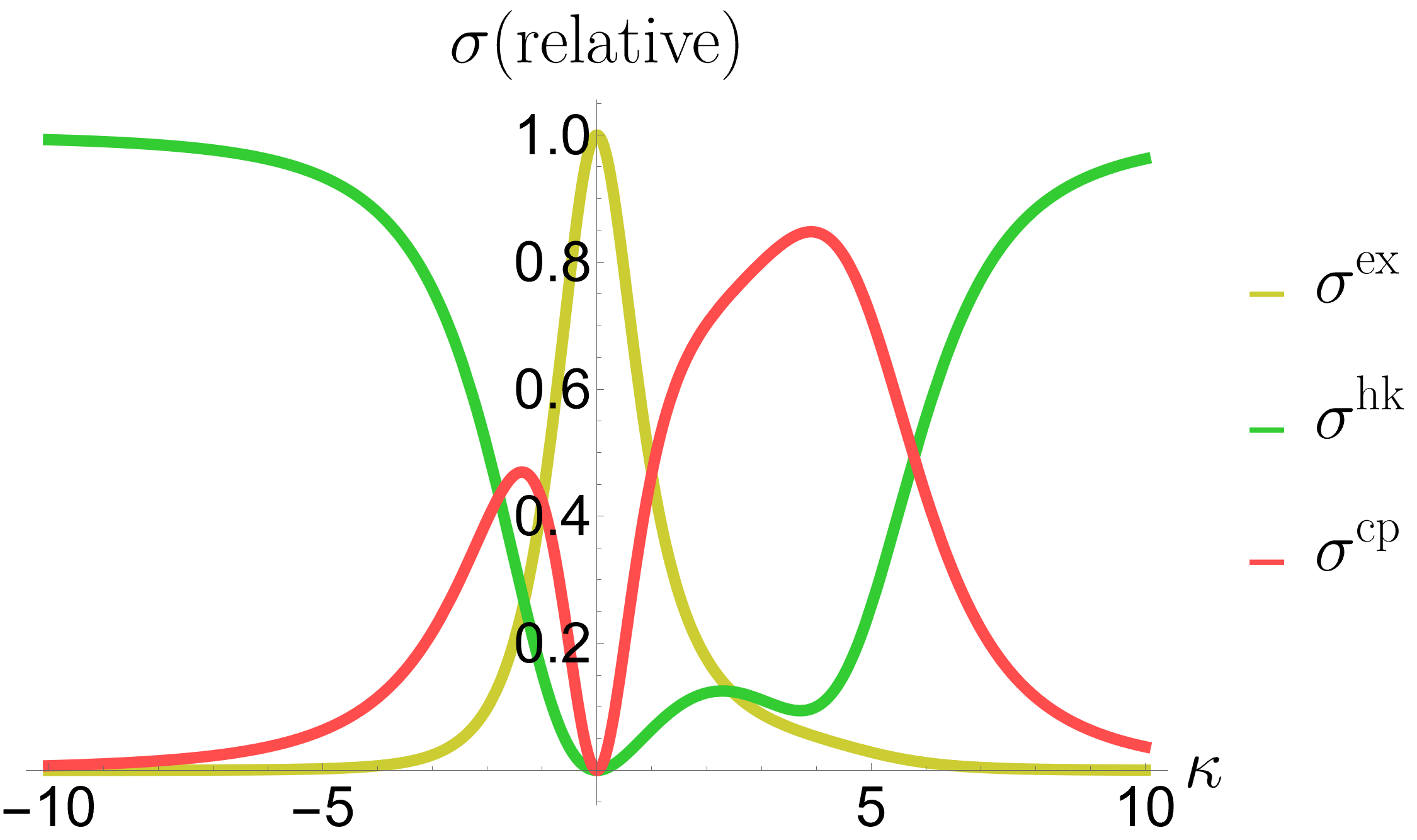}
\caption{The entropy production rate and its decomposition for the model illustrated in Fig.~\ref{fig-example} as a function of the strength $\kappa$ of the nonconservative driving.
Top panel: The entropy production rate (black), the excess part (yellow/light gray), housekeeping part (green/medium gray) and coupling part (red/dark gray).
Bottom panel: The excess part (yellow/light gray), housekeeping part (green/medium gray) and coupling part (red/dark gray) relative to the overall entropy production rate.
The remaining parameters are $\mu = 1$, $k = 1$ and $T = 0.5$; the potential moves around the origin in a circle with radius $a = 3$ with frequency $\omega = 4$.  \label{fig-example-results}}
\end{figure}
To demonstrate the decomposition \eqref{decomposition} explicitly, we consider the following example.
A Brownian particle is trapped in a two-dimensional parabolic potential $U_t(\bm{x}) = k \Vert \bm{x} - \bm{a}_t \Vert^2/2$ whose center position $\bm{a}_t$ changes with time.
In addition, the particle is driven by the nonconservative force $\bm{F}^\text{nc}(\bm{x}) = \kappa (-x_2,x_1)$, which corresponds to a torque driving the particle in counter-clockwise direction (for $\kappa > 0$) around the origin of the $x_1$-$x_2$-plane.
This model is an extension of the so-called Brownian gyrator \cite{Fil07,Kwo11} to include a time-dependent trapping potential.
The setup is illustrated in Fig.~\ref{fig-example}.
The Fokker-Planck equation \eqref{fokkerplanck} for this system reads
\begin{align}
\partial_t &p_t(x_1,x_2) = \mu \Big( \partial_{x_1} \big( k (x_1 - a_{1,t}) + \kappa x_2 - T \partial_{x_1} \big) \\
&+ \partial_{x_2} \big( k (x_2 - a_{2,t}) - \kappa x_1 - T \partial_{x_2} \big) \Big) p_t(x_1,x_2) .
\end{align}
Since the forces are linear, provided that the initial state of the system is Gaussian, $p_t(\bm{x})$ is Gaussian at any time and characterized by its mean and covariance matrix, see Appendix \ref{app-gaussian}.
Moreover, the time evolution of the mean and covariance matrix decouple, and from \eqref{moment-equations}, the mean evolves according to
\begin{subequations}
\begin{align}
d_t \av{x_1}_t &= - \mu \big( k (\av{x_1}_t - a_{1,t}) + \kappa \av{x_2}_t \big), \\
d_t \av{x_2}_t &= - \mu \big( k (\av{x_2}_t - a_{2,t}) - \kappa \av{x_1}_t \big) .
\end{align} \label{mean-evolution}%
\end{subequations}
Since the time-dependent driving does not enter the equations for the covariance matrix, we focus on the case where the covariance matrix has relaxed to its steady state value $\text{Var}(x_1) = \text{Var}(x_2) = T/k$, $\text{Cov}(x_1,x_2) = 0$,
\begin{align}
p_t(x_1,x_2) = \frac{k}{2 \pi T} \exp\bigg( - \frac{k \Vert \bm{x} - \av{\bm{x}}_t \Vert^2}{2 T} \bigg) .
\end{align}
Further, as illustrated in Fig.~\ref{fig-example}, we focus on the case where the motion of the trap is periodic around the origin. 
Then, the system will settle into a periodic state at long times.
If the trap moves around the origin in a circle of radius $a$ with angular frequency $\omega$, we obtain for the mean,
\begin{align}
\begin{pmatrix} \av{x_1}_t \\[1ex] \av{x_2}_t \end{pmatrix} &= \frac{a k}{k^2 + \big(\kappa - \frac{\omega}{\mu} \big)^2} \\ 
&\qquad \times \begin{pmatrix} k \cos(\omega t) - \big(\kappa - \frac{\omega}{\mu} \big) \sin(\omega t) \\[1ex] k \sin(\omega t) + \big(\kappa - \frac{\omega}{\mu} \big) \cos(\omega t) \end{pmatrix} \n ,
\end{align}
while its instantaneous steady state value is given by
\begin{align}
\begin{pmatrix} \av{x_1}^\text{st}_t \\[1ex] \av{x_2}^\text{st}_t \end{pmatrix} &= \frac{a k}{k^2 + \kappa^2} \\ 
&\qquad \times \begin{pmatrix} k \cos(\omega t) - \kappa \sin(\omega t) \\[1ex] k \sin(\omega t) + \kappa \cos(\omega t) \end{pmatrix} \n .
\end{align}
In terms of these expressions, the local mean velocity and its instantaneous steady state value are
\begin{subequations}
\begin{align}
\bm{\nu}_t(\bm{x}) &= - \mu k \big( \av{\bm{x}}_t - \bm{a}_t \big) + \mu \bm{F}^\text{nc}(\bm{x}) \\
\bm{\nu}^\text{st}_t(\bm{x}) &= - \mu k \big( \av{\bm{x}}^\text{st}_t - \bm{a}_t \big) + \mu \bm{F}^\text{nc}(\bm{x}) .
\end{align}
\end{subequations}
Since the minimum entropy production dynamics for this time evolution is a particle in a time-dependent parabolic potential without the nonconservative force, the corresponding local mean velocity is given by
\begin{align}
\bm{\nu}_t^*(\bm{x}) = d_t \av{\bm{x}}_t \label{example-minent},
\end{align}
With these results, it is straightforward to calculate the explicit expressions for the decomposition \eqref{decomposition} of the entropy production rate; the result is
\begin{subequations}
\begin{align}
\sigma_t^\text{ex} &= \frac{a^2 \omega^2 k^4}{\mu T (k^2 + \kappa^2) \big(k^2 + \big(\kappa - \frac{\omega}{\mu} \big)^2 \big)}, \\
\sigma_t^\text{hk} &= \frac{2 \mu \kappa^2}{k}, \\
\sigma_t^\text{cp} &= \frac{a^2 \omega^2 k^2 \kappa^2}{\mu T (k^2 + \kappa^2) \big(k^2 + \big(\kappa - \frac{\omega}{\mu} \big)^2 \big)}, \\
\sigma_t &= \frac{2 \mu \kappa^2}{k} + \frac{a^2 \omega^2 k^2}{\mu T \big(k^2 + \big(\kappa - \frac{\omega}{\mu} \big)^2 \big)} ,
\end{align} \label{decomposition-example}%
\end{subequations}
see also \eqref{decomposition-gaussian}.
The behavior of the entropy production rate and its decomposition is shown in Fig.~\ref{fig-example-results}.
We see that the entropy production rate exhibits a marked peak (top panel) when rotation frequency due to the nonconservative force $\mu \kappa$ and the rotation frequency of the potential $\omega$ have the same value.
For these parameters, the nonconservative and the time-dependent driving forces align to produce a greatly enhanced motion of the particle, leading to increased dissipation.
By considering the decomposition \eqref{decomposition} we see that this peak is almost entirely due to a corresponding peak in the coupling entropy production rate, confirming its interpretation as quantifying the interaction between time-dependent and nonconservative driving.
This becomes even more apparent when considering the size of the three contributions relative to the total entropy production rate (bottom panel).
As expected, for small nonconservative forces, the excess part dominates, while for large nonconservative forces, the housekeeping part becomes dominant.
But in the intermediate regime, where the nonconservative and time-dependent forces are comparable, the coupling part yields the dominant contribution.

\begin{figure}
\includegraphics[width=.47\textwidth]{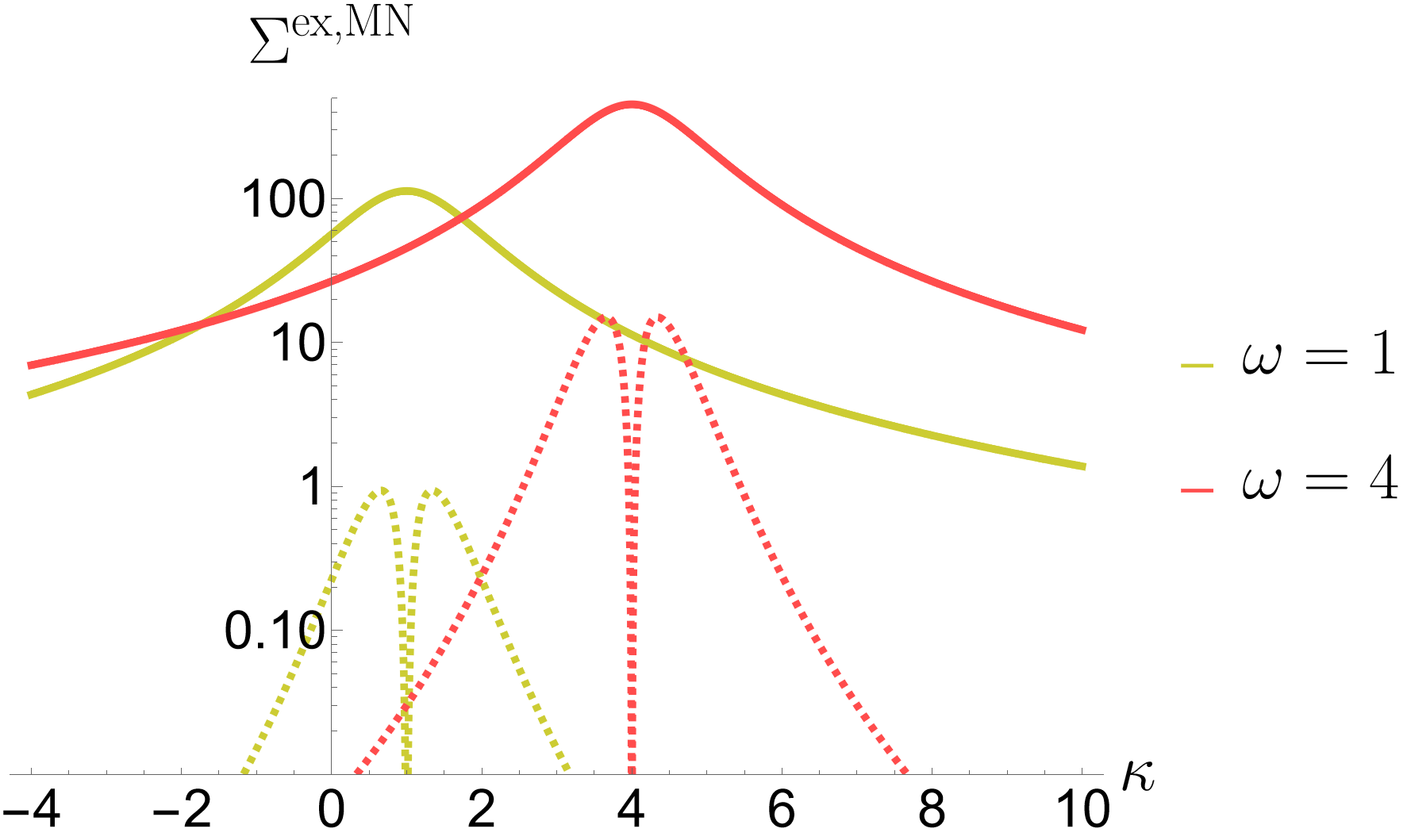}\\
\includegraphics[width=.47\textwidth]{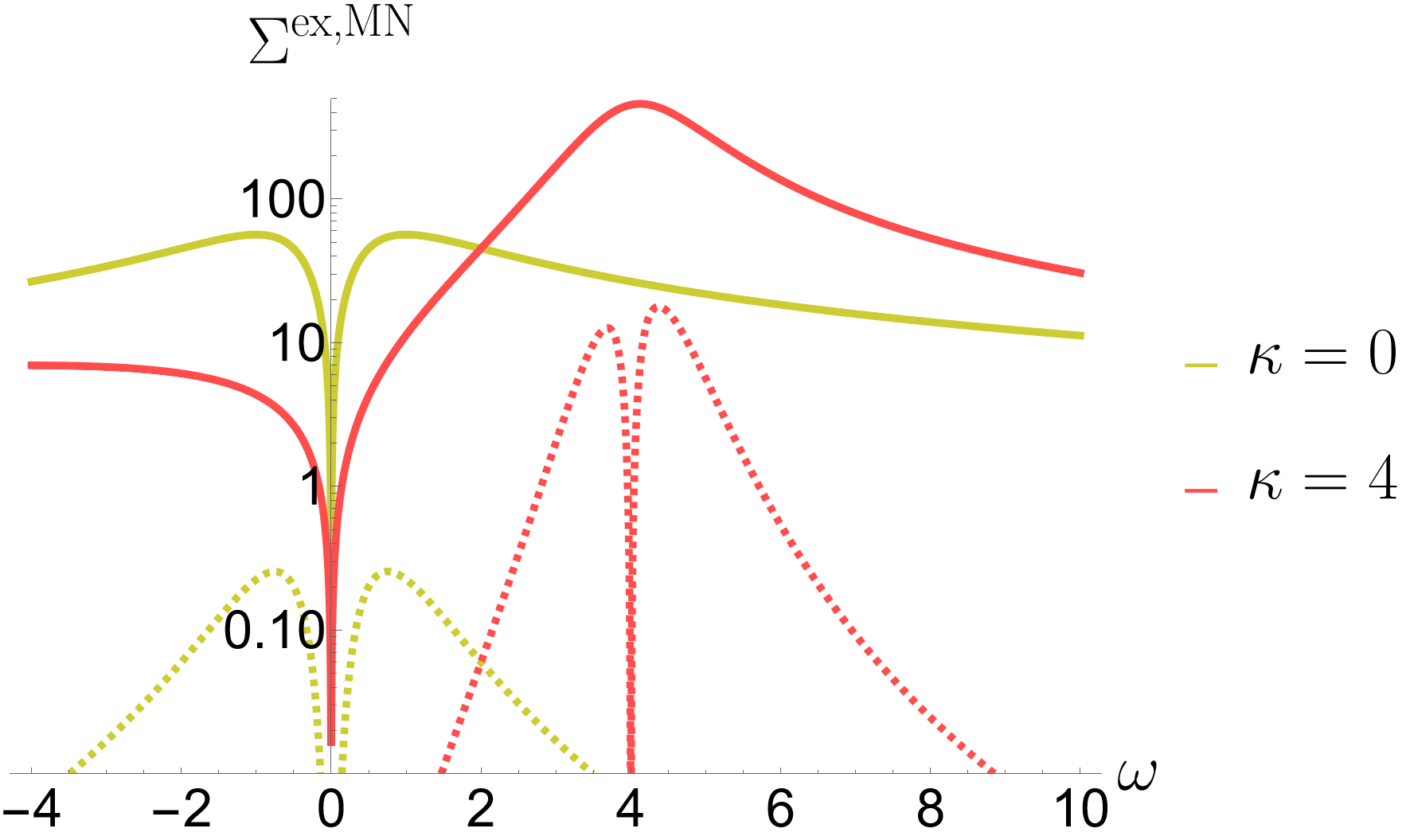}
\caption{Comparison between the lower bound from the finite-time TUR \eqref{excess-TUR-example} on the excess entropy production per period (dashed) and the actual value (solid). The top panel shows the results as a function of the magnitude of the nonconservative force for two different driving frequencies, the bottom panel as a function of the driving frequency for two different nonconservative forces (positive/negative frequency corresponds to a counterclockwise/clockwise motion of the trap. The remaining parameter values are $\mu = 1$, $k = 1$, $T = 0.5$ and $a = 3$. \label{fig-example-tur}}
\end{figure}
While, as shown above, we can explicitly calculate the different contributions to the entropy production in this example, in practice, we may not know the precise values of all the parameters in the system.
In this case, the variational expressions derived in Section \ref{sec-variational} and the lower bounds derived in Section \ref{sec-tur} can be used to estimate some contributions from measured observables.
In particular, we recall \eqref{variational-MN}, which, for the present case, can be written as \cite{Dec21b},
\begin{subequations}
\begin{align}
    \sigma_t^\text{ex,MN} &= \sup_{\eta} \bigg( \frac{\big(d_t \av{\eta}_t \big)^2}{\mu T \av{\Vert\grad \eta \Vert^2}_t} \bigg), \\
    \sigma_t^\text{hk,MN} &= \inf_{\eta} \bigg( \frac{\mu}{T} \Av{ \Vert \bm{F}^\text{nc} - T \grad \eta \Vert^2}_t \bigg) ,
\end{align}
\end{subequations}
where the maximization/minimization is performed over all scalar functions $\eta(\bm{x})$.
Here, we used that the space $V_1^\text{MN}$ in \eqref{variational-MN} is the space of all gradient fields.
Thus, if we know the bare diffusivity $\mu T$, we can calculate the MN excess entropy production rate by maximizing the rate of change in the average of a scalar observable relative to the magnitude of its gradient.
Importantly, both the average rate of change and the magnitude of the gradient of some arbitrary function $\eta(\bm{x})$ can be computed from measured trajectories of the system.
Similarly, we can calculate the MN housekeeping entropy production rate from the variational formula if, in addition, we know the nonconservative force $\bm{F}^\text{nc}(\bm{x})$.
However, even it is unfeasible to perform the optimization over all possible scalar functions, we can always obtain bounds by restricting ourselves to a particular set of functions.
In this case, since the system is linear, a reasonable ansatz for $\eta(\bm{x})$ is
\begin{align}
    \eta(\bm{x}) = \alpha_1 x_1 + \alpha_2 x_2 = \bm{\alpha} \cdot \bm{x},
\end{align}
with some vector $\bm{\alpha}$.
We then obtain the bounds
\begin{subequations}
\begin{align}
    \sigma_t^\text{ex,MN} &\geq \sup_{\bm{\alpha}} \bigg( \frac{\big( \bm{\alpha} \cdot d_t \av{\bm{x}}_t \big)^2}{\mu T \Vert \bm{\alpha} \Vert^2} \bigg), \\
    \sigma_t^\text{hk,MN} &\leq \inf_{\bm{\alpha}} \bigg( \frac{\mu}{T} \Av{ \Vert \bm{F}^\text{nc} - T \bm{\alpha} \Vert^2}_t \bigg) .
\end{align}
\end{subequations}
The optimization can now be performed explicitly, and we obtain $\bm{\alpha} = d_t \av{\bm{x}}_t$ in the first line and $\bm{\alpha} = \av{\bm{F}^\text{nc}}_t/T$ in the second line, which leads to
\begin{subequations}
\begin{align}
    \sigma_t^\text{ex,MN} &\geq \frac{\Vert d_t \av{\bm{x}}_t \Vert^2}{\mu T} \\
    \sigma_t^\text{hk,MN} &\leq \frac{\mu}{T} \Av{ \Vert \bm{F}^\text{nc} - \av{\bm{F}^\text{nc}}_t \Vert^2}_t .
\end{align} \label{ex-hk-bound}%
\end{subequations}
We remark that these bounds are completely general; the first one corresponds to the short-time TUR \eqref{short-time-TUR-excess-MN} for the weighting function $\bm{w}_t(\bm{x}) = d_t \av{\bm{x}}_t$.
Comparing this to \eqref{decomposition-example}, we find that, in the present example, both inequalities are actually equalities, which shows that these bounds can be tight.
Equation~(\ref{ex-hk-bound}) provides useful and general bounds on the excess and housekeeping entropy, but we still need to know the temperature, the mobility and the nonconservative force to evaluate them.
However, for the MN excess entropy, we can use the finite-time TUR \eqref{excess-TUR} to obtain a lower bound that can be evaluated directly in terms of measured quantities.
Suppose that we measure the instantaneous squared distance of the particle from the minimum of the potential,
\begin{align}
    J_\tau = \Vert \bm{x}(r \tau) - \bm{a}_{r \tau} \Vert^2 ,
\end{align}
where $r \in [0,1]$ denotes some fixed point along the driving protocol.
Then, from \eqref{excess-TUR}, we have
\begin{align}
    \frac{\big(\tau d_\tau \av{J_\tau} \big)^2}{\text{Var}(J_\tau)} \leq \frac{1}{2} \Sigma_\tau^\text{ex,MN} \label{excess-TUR-example} .
\end{align}
Note that, since the driving is periodic, a changing the driving period $\tau$ is equivalent to changing the driving frequency $\omega = 2 \pi/\tau$, and thus $\tau d_\tau = - \omega d_\omega$.
Thus, we can obtain a lower bound on the excess entropy by examining the dependence of the lag between the minimum position of the trap and the actual position of the particle on the speed of driving.
Since the the probability density is Gaussian, all quantities can be evaluated explicitly, allowing us to directly assess the tightness of the bound.
We obtain,
\begin{align}
    \av{J_\tau} &= \Vert \av{\bm{x}}_{r\tau} - \bm{a}_{r\tau} \Vert^2 + \frac{2 T}{k}, \\
    \text{Var}(J_\tau) &= \frac{4 T}{k} \Big( \Vert \av{\bm{x}}_{r\tau} - \bm{a}_{r\tau} \Vert^2 + \frac{T}{k} \Big) \n .
\end{align}
Since the dynamics are radially symmetric, the lag is independent of the point $r$ along the protocol,
\begin{align}
\Vert \av{\bm{x}}_{r\tau} - \bm{a}_{r\tau} \Vert^2 = a^2 \frac{\big(\kappa - \frac{\omega}{\mu} \big)^2}{k^2 + \big(\kappa - \frac{\omega}{\mu} \big)^2} .
\end{align}
The results are shown in Fig.~\ref{fig-example-tur}.
We see that this choice of the observable does not capture the magnitude of the excess entropy production; the lower bound is at most $0.04$ of the true value.
In order to improve upon this, we would have to find an observable that is more sensitive to changes in the driving frequency.
However, we also see that the lower bound captures several qualitative features of the excess entropy, including its maximum due to the coupling contribution (in terms of \eqref{decomposition}, the MN excess entropy contains both the excess and the coupling term) when the driving frequency matches the magnitude of the nonconservative force.

\subsection{Periodic change in the stiffness of a parabolic trap}
\begin{figure*}
    \centering
    \includegraphics[width=.47\textwidth]{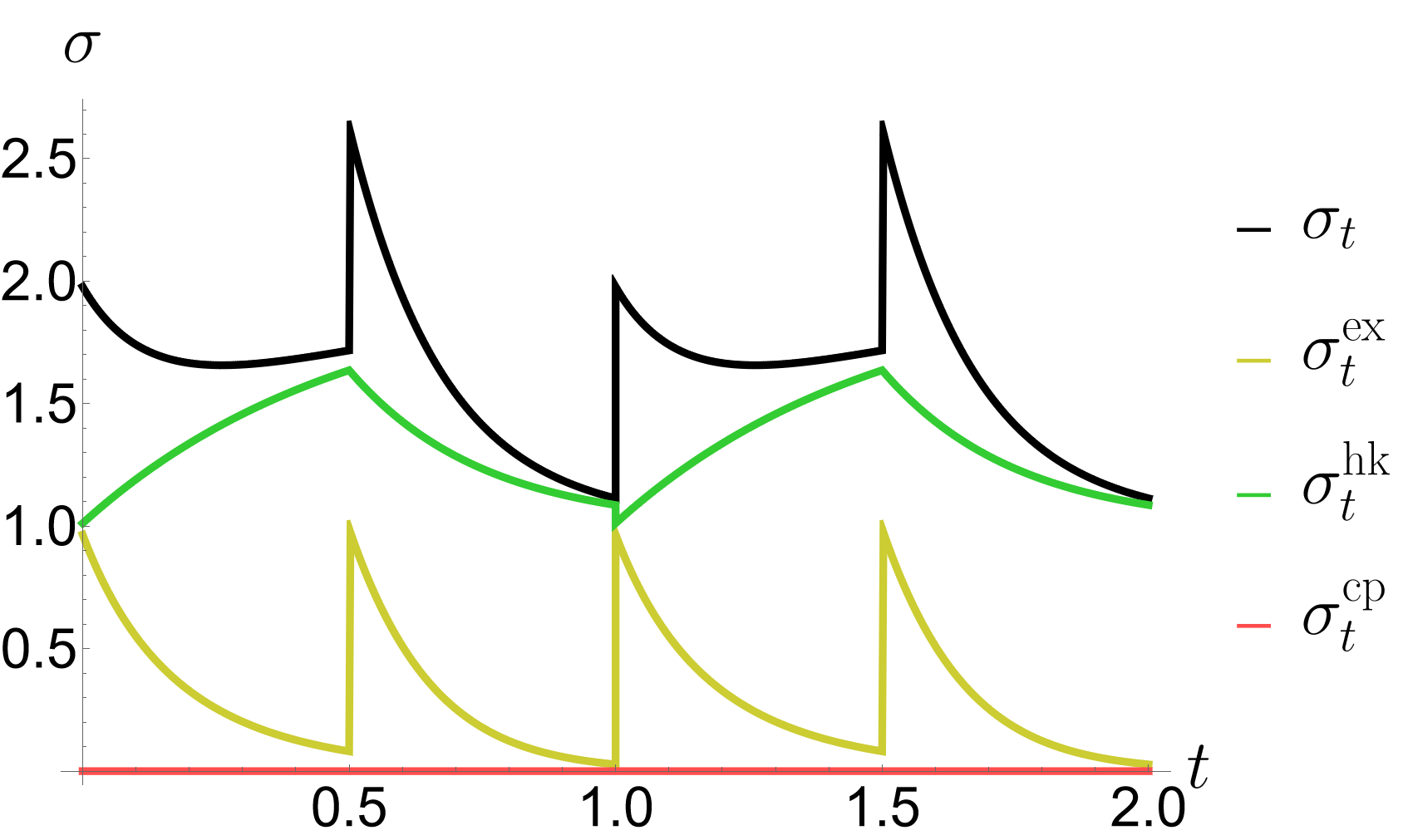}
    \includegraphics[width=.47\textwidth]{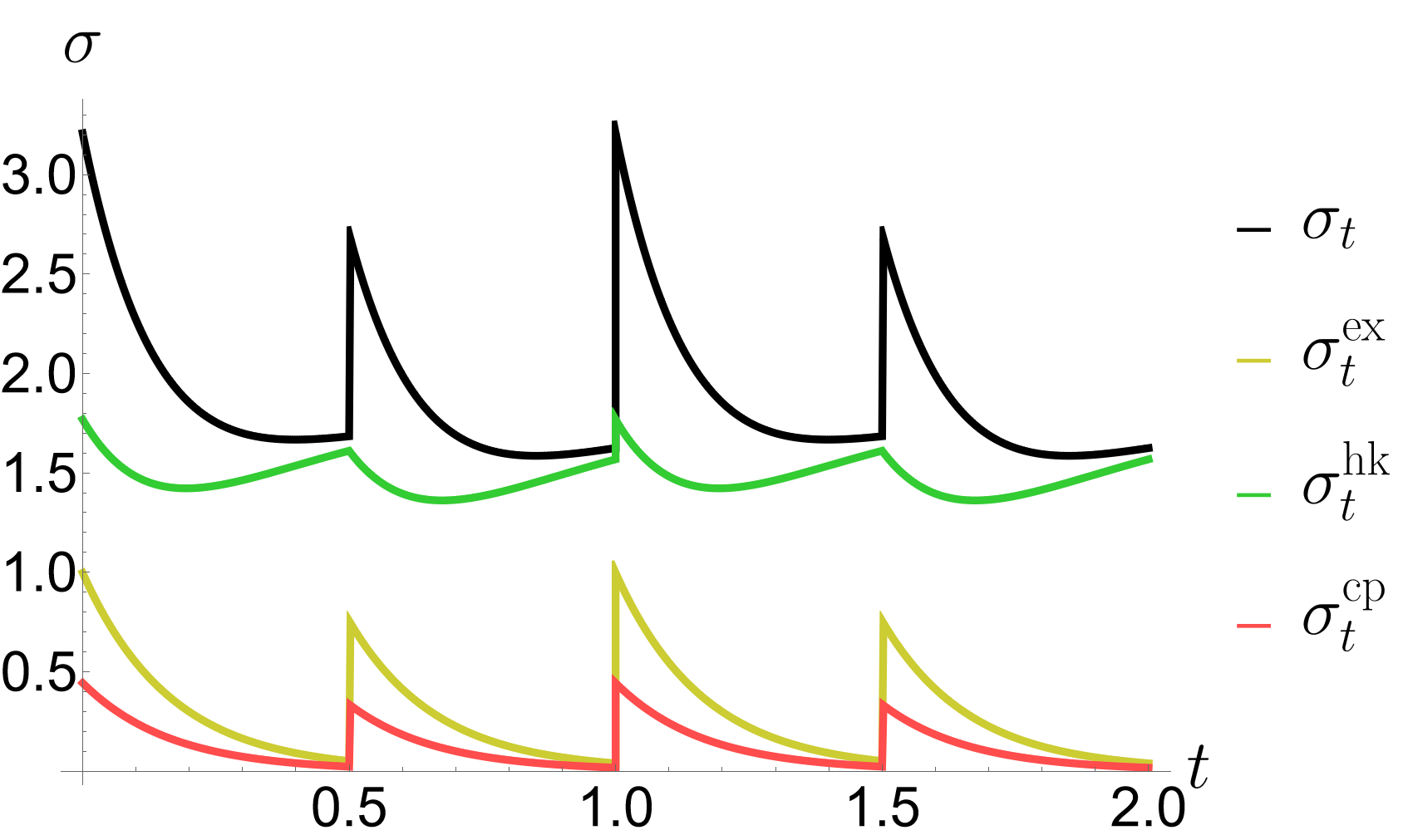}\\
    \includegraphics[width=.47\textwidth]{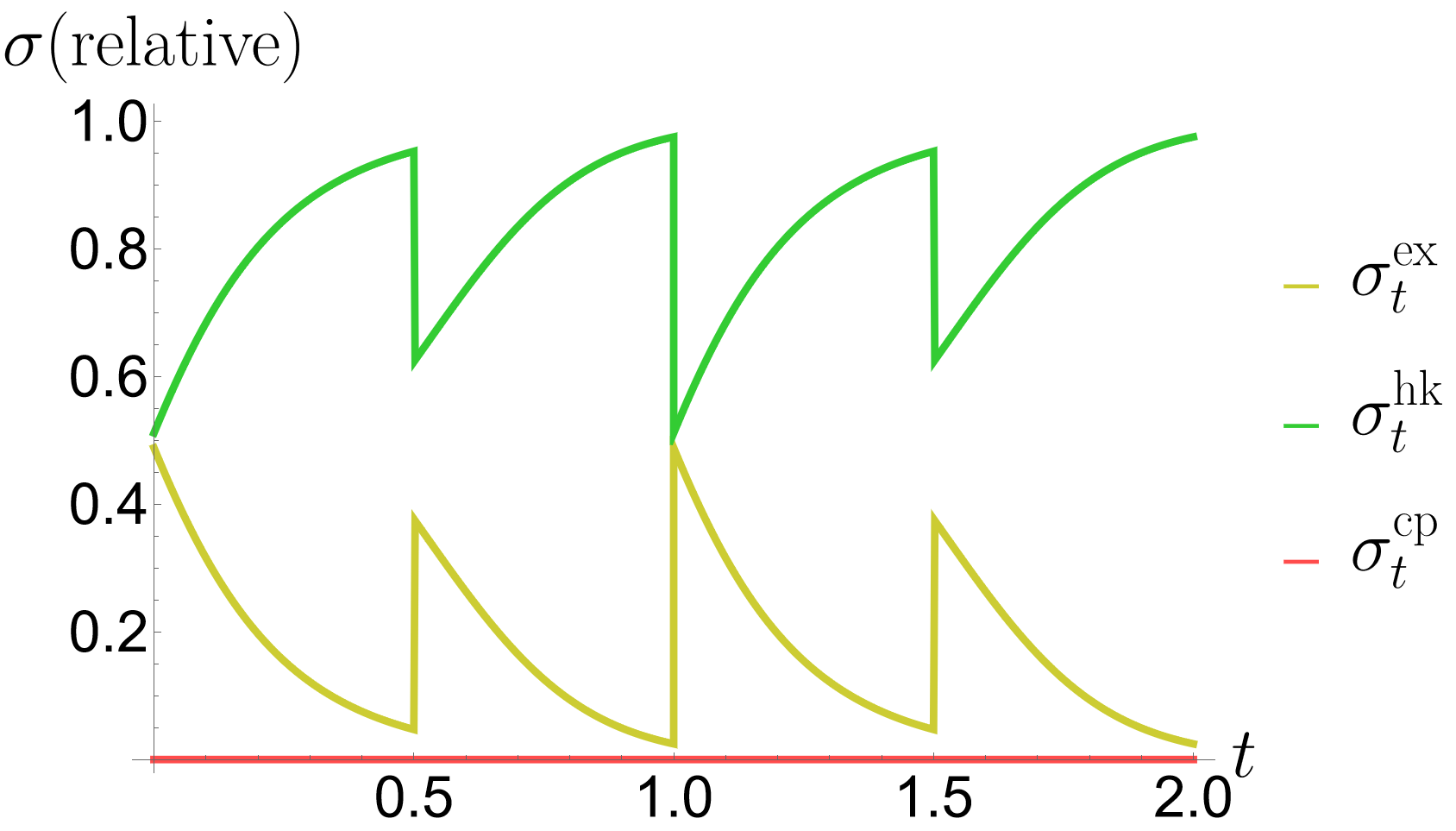}
    \includegraphics[width=.47\textwidth]{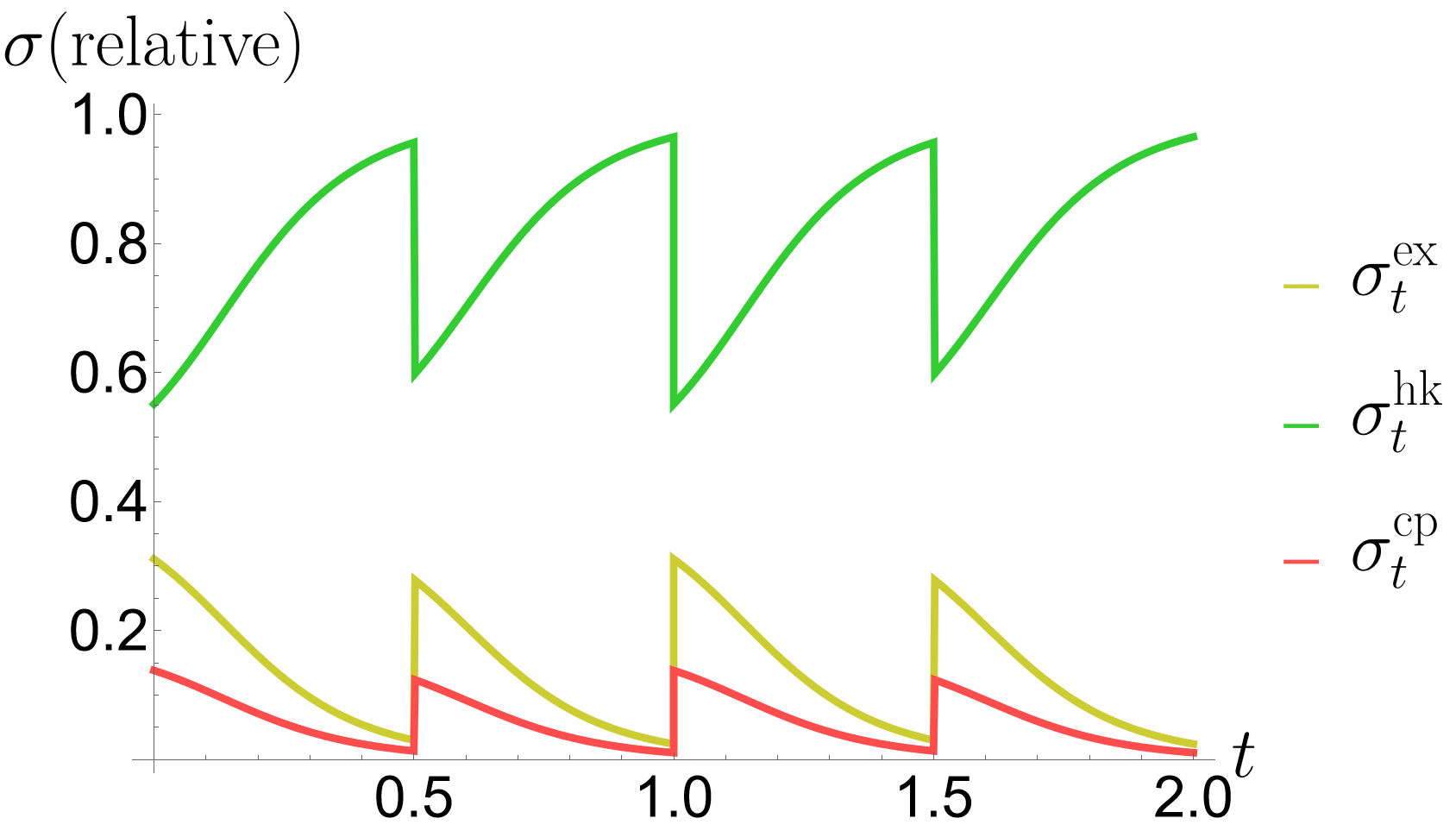}
    \caption{The entropy production and its excess, housekeeping and coupling part \eqref{decomposition} for changing the trap stiffness in the presence of a nonconservative force. The left column corresponds to a radially symmetric trap with changing overall stiffness, $\bm{K}_t$ in \eqref{trap-matrix}, the right column to a nonsymmetric trap with changing symmetry, $\tilde{\bm{K}}_t$ in  \eqref{trap-matrix}. The top row shows the absolute magnitude of the individual parts, the bottom row shows the magnitude of the parts relative to the total entropy production rate. The parameters are $\mu = 1$, $T = 1$, $\kappa = 1$, $k_0 = 1$, $k_1 = 2$ and $\tau = 0.5$.}
    \label{fig-example-trap}
\end{figure*}

Finally, we construct an example where the coupling part vanishes while both the excess and housekeeping parts are finite.
As shown in Section \ref{sec-coupling}, a vanishing coupling entropy is equivalent to the condition \eqref{zero-coupling-orthogonal}.
For the above example, this implies
\begin{align}
\big( \bm{F}^\text{nc}(\bm{x}) - \av{\bm{F}^\text{nc}}_t \big) \cdot d_t \av{\bm{x}}_t = 0 \label{coupling-zero-condition} .
\end{align}
This expression remains true as long as the trapping potential is parabolic, $U_t(\bm{x}) = k \Vert \bm{x} - \bm{a}_t \Vert^2/2$, and the probability density is given by the Gaussian
\begin{align}
p_t(\bm{x}) = \bigg(\frac{k}{2 \pi T} \bigg)^{\frac{d}{2}} e^{-\frac{k}{2 T}\Vert \bm{x} - \av{\bm{x}}_t \Vert^2}.
\end{align}
In two dimensions, \eqref{coupling-zero-condition} cannot be satisfied in a nontrivial way:
If the nonconservative force is orthogonal to a constant (with respect to the coordinates) vector field, then it has to be of the form $\bm{F}^\text{nc}(\bm{x}) = g(\bm{x}) \bm{u}_t$, where $\bm{u}_t$ is the unit vector of the direction orthogonal to $d_t \av{\bm{x}}_t$.
However, such a force can always be written as a gradient and is thus conservative.
Thus, for the two-dimensional example discussed above, the only possibilities for a vanishing coupling entropy are $\bm{F}^\text{nc} = 0$ or $d_t \av{\bm{x}}_t = 0$, which implies that the housekeeping, respectively excess, entropy vanish as well.
In three dimensions, on the other hand, there is a simple way of satisfying \eqref{coupling-zero-condition}:
If we apply a nonconservative force as above in the $x_1-x_2$-plane and a time-dependent driving in the $x_3$-direction, then the effects of the two types of driving are independent of each other.
In \eqref{zero-coupling-orthogonal}, $\bm{\nu}_t^*(\bm{x})$ is only has a $x_3$-component, while $\bm{\nu}_t(\bm{x}) - \bm{\nu}_t^*(\bm{x})$ only has $x_1$ and $x_2$ components, so they are trivially orthogonal.
In two dimensions, we can also obtain a nontrivial example satisfying \eqref{coupling-zero-condition}, if we allow the width of the distribution to depend on time,
\begin{align}
p_t(x_1,x_2) = \frac{1}{2 \pi \Xi_t} e^{-\frac{1}{2 \Xi_t} \Vert \bm{x} \Vert^2} ,
\end{align}
for example, via a time-dependent trapping strength $U_t(\bm{x}) = \frac{k_t}{2} \Vert \bm{x} \Vert^2$.
In this case, since the probability density is radially symmetric, the flows $\bm{\nu}_t^*(\bm{x})$ contributing to the time evolution only have a radial component, while the nonconservative force $\bm{F}^\text{nc}(\bm{x}) = \kappa (-x_2,x_1)$ and $\bm{\nu}_t(\bm{x}) - \bm{\nu}_t^*(\bm{x})$ only has a tangential component, thus satisfying \eqref{zero-coupling-orthogonal}.
More generally, let us consider a potential of the form $U_t(\bm{x}) = \frac{1}{2} \bm{x} \cdot \bm{K}_t \bm{x}$, where $\bm{K}_t$ is a symmetric matrix.
Such a potential is generally not radially symmetric and the matrix $\bm{A}_t$ in \eqref{gaussian-langevin} can be written as
\begin{align}
    \bm{A}_t = \mu \begin{pmatrix} K_{11,t} & K_{12,t} + \kappa \\ K_{12,t} - \kappa & K_{22,t} \end{pmatrix} .
\end{align}
For simplicity, we focus on the case where $\bm{a}_t = 0$ (so that the distribution remains centered at $\bm{x} = 0$) and the matrix $\bm{K}_t$ is instantaneously and periodically changed between two values $\bm{K}_0$ and $\bm{K}_1$ with time $\tau$ between subsequent changes.
In this case, the dynamics consists of a sequence of relaxation processes and we can formally solve \eqref{moment-equations},
\begin{align}
    \bm{C}_t = \left\lbrace \begin{array}{l} 2 \mu T \int_0^t ds \ e^{-\bm{A}_0 s} e^{-\bm{A}_0^\text{T} s} + e^{-\bm{A}_0 t} \bm{C}_0 e^{\bm{A}_0^\text{T} t} \\
    \hspace{2cm} \text{for} \; 0 \leq t \leq \tau \\
    2 \mu T \int_0^{t-\tau} ds \ e^{-\bm{A}_1 s} e^{-\bm{A}_1^\text{T} s} \\
    \hspace{.5cm} + e^{-\bm{A}_1 (t-\tau)} \bm{C}_\tau e^{\bm{A}_1^\text{T} (t-\tau)} \quad \text{for} \; \tau < t \leq 2 \tau,
    \end{array} \right.
\end{align}
and the constant $\bm{C}_0$ is determined by the condition $\bm{C}_{2\tau} = \bm{C}_0$.
Note that the matrices $\bm{A}_0$ and $\bm{A}_1$ do not commute with each other and their transposes.
In principle, we can obtain an explicit expression in terms of the parameters of the model, however, this expression is already so complicated as to be of little practical use, and we therefore evaluate the covariance matrix numerically and then use \eqref{decomposition-gaussian} to compute the respective contriubtions to the entropy production rate.
We consider the two specific cases,
\begin{subequations}
\begin{align}
    \bm{K}_0 &= \begin{pmatrix} k_0 & 0 \\ 0 & k_0 \end{pmatrix}, \qquad \bm{K}_1 = \begin{pmatrix} k_1 & 0 \\ 0 & k_1 \end{pmatrix} \\
    \tilde{\bm{K}}_0 &= \begin{pmatrix} k_0 & 0 \\ 0 & k_1 \end{pmatrix}, \qquad \tilde{\bm{K}}_1 = \begin{pmatrix} k_1 & 0 \\ 0 & k_0 \end{pmatrix} .
\end{align} \label{trap-matrix}%
\end{subequations}
The first case, $\bm{K}_t$, corresponds to a radially symmetric trap, whose stiffness is periodically changed between two values.
By contrast, in the second case, the trap is not radially symmetric (for $k_1 > k_0$, the particle is more strongly confined in the $x_2$-direction), and the symmetry is changed periodically.
The results for the individual contributions to the entropy production rate are shown in Fig.~\ref{fig-example-trap}.
In both cases, we first note that the total entropy production rate can behave in a nonmonotonic manner even during a relaxation process.
The excess entropy production rate, on the other hand, is maximal directly after changing the parameters and then monotonously decreases.
Most importantly, we see that, as argued above, the coupling entropy production rate vanishes when the trapping potential is radially symmetric; in this case, the change in the probability density only occurs in the radial direction and is thus orthogonal to the steady state flow due to the nonconservative force.
By contrast, when the radial symmetry of the potential is broken, we observe a finite coupling entropy production, which qualitatively behaves in a similar way as the excess entropy.
Thus, as we argued in Section \ref{sec-coupling}, the coupling entropy production rate is nonzero in the generic case, and its absence implies the existence of special symmetries in the system.

\section{Discussion} \label{sec-discussion}
In general, when a system is affected by both time-dependent and nonconservative forces, their effects can combine in a nontrivial way.
As we demonstrated using the solvable models discussed in Section \ref{sec-example}, the contribution of the coupling entropy to the total entropy production can be significant in this case.
By contrast, it vanishes when the two types of driving act independently on different degrees of freedom of the system.
This suggest that we can use the coupling term to quantify the interdependence between these two qualitatively different types of driving.

One possible application is pumping \cite{Che08,Rah08}:
According to the second law of thermodynamics, it is not possible to extract work from a single heat bath using a periodic process.
That is, if we change the potential $U_t(\bm{x})$ in a periodic manner, then the work $W = \int_0^\tau dt \ \partial_t U_t(\bm{x}(t))$ done on the system will always be positive on average.
However, this is no longer true if the system is driven out of equilibrium by a nonconservative force.
In this case, we may in principle extract work through a periodic process; the work of course being supplied by the nonconservative force.
Intuitively, it is clear that, in order to extract work, we should perform the periodic operation on the degrees of freedom that are affected by the nonconservative force. 
That is, we require a coupling between the nonconservative force and the time-dependent protocol, which suggests that the extracted work should be related to the coupling entropy.

The discussion in Section \ref{sec-coupling} indicates that we may decompose the local mean velocity $\bm{\nu}_t(\bm{x})$ into three orthogonal components, independent of the dimensionality of the system.
This may seem surprising, however, we stress that this orthogonality is defined with respect to the inner product \eqref{inner-product}, which permits three orthogonal components even in one- or two-dimensional systems.
In general, we can consider different types of orthogonality, for example with respect to the usual inner product in $\mathbb{R}^d$, or with respect to some function space.
The orthogonality considered here is a combination of both, which involves the inner product between vector fields and averaging with respect to the probability density describing the current state of the system.
This implies that, generically, a system driven by a time-dependent and a nonconservative force has at least three degrees of freedom, corresponding to the three components of the local mean velocity.
In general, these degrees of freedom are not simply related to the control parameters of the system, since, for example, changing the magnitude of the nonconservative force modifies all three contributions to the entropy production.

A challenging yet interesting problem is the generalization of the results in this article to underdamped Langevin dynamics.
While a Hatano-Sasa-type decomposition has been developed for the underdamped case \cite{Spi12b,Lee13,Lah14}, an extension of the Maes-Neto{\v{c}}n{\`y} formalism seems more challenging, since minimizing the entropy production rate in underdamped dynamics does not yield a unique potential force \cite{Mur14,Mur14b}.
However, if such a generalization can be developed, the corresponding coupling entropy may prove useful in studying tracer particles in an active bath \cite{Mae20}.
In this situation, the particles constituting the active bath constantly dissipate energy, which may be interpreted as housekeeping entropy.
By contrast, performing a time-dependent operation on the tracer particle leads to a non-vanishing excess entropy, while the coupling entropy could quantify the nontrivial interactions between the dissipation of the bath and the tracer.

\begin{acknowledgments}
We thank Takuya Kamijima and Takahiro Sagawa for valuable discussions on thermodynamic uncertainty relations. A.~D.~is supported by JSPS KAKENHI (Grant No. 19H05795, and 22K13974). S.~S.~is supported by JSPS KAKENHI (Grant No. 19H05795, 20K20425, and 22H01144). S.~I.~is supported by JSPS KAKENHI (Grant No. 19H05796, and 21H01560), JST Presto (Grant No. JPMJPR18M2) and UTEC-UTokyo FSI Research Grant Program.
\end{acknowledgments}

\appendix

\onecolumngrid

\section{Decomposition of entropy production for Gaussian processes} \label{app-gaussian}
For general systems, while the decomposition \eqref{decomposition} always exists, its explicit computation is challenging and requires the time-dependent probability density $p_t(\bm{x})$, as well as the instantaneous steady state $p_t^\text{st}(\bm{x})$ and instantaneous canonical density $p_t^\text{can}(\bm{x})$.
The solution can be made more explicit for systems that are driven by forces which are linear in the degrees of freedom.
The general equation of motion is of the form
\begin{align}
    \dot{\bm{x}}(t) = -\mu \big(\bm{A}_t \bm{x}(t) - \bm{a}_t \big) + \sqrt{2 \mu T} \bm{\xi}(t), \label{gaussian-langevin}
\end{align}
with some matrix $\bm{A}_t$ and vector $\bm{a}_t$.
The existence of a steady state requires that all eigenvalues of the matrix $\bm{A}_t$ have a positive real part; however, they can be complex, since the matrix $\bm{A}_t$ is generally not symmetric.
Provided that the initial probability density is Gaussian, the time-dependent solution is also Gaussian for all times and can be written as
\begin{align}
    p_t(\bm{x}) &= \sqrt{\frac{1}{(2\pi)^d \det(\bm{C}_t)}} \exp \bigg(- \frac{1}{2} \big(\bm{x} - \bm{m}_t \big) \cdot \bm{C}_t^{-1} \big(\bm{x} - \bm{m}_t \big) \bigg) .
\end{align}
The time-dependent mean $\bm{m}_t$ and the covariance matrix $\bm{C}_t$ (which is symmetric and positive definite) satisfy the equations of motion
\begin{subequations}
\begin{align}
    d_t \bm{m}_t &= - \mu \big(\bm{A}_t \bm{m}_t - \bm{a}_t \big), \\
    d_t \bm{C}_t &= - \mu \big( \bm{A}_t \bm{C}_t + \bm{C}_t \bm{A}_t^\text{T} \big) + 2 \mu T \bm{I} ,
\end{align} \label{moment-equations}%
\end{subequations}
where the superscript $T$ denotes transposition and $\bm{I}$ the $d\times d$ identity matrix.
Similarly, the instantaneous steady state is Gaussian, with its mean and covariance matrix determined by
\begin{subequations}
\begin{align}
   0 &= - \bm{A}_t \bm{m}^\text{st}_t + \bm{a}_t, \\
   0 &= - \big(\bm{A}_t \bm{C}^\text{st}_t + \bm{C}^\text{st}_t \bm{A}_t^\text{T} \big) + 2 T \bm{I} .
\end{align} \label{moment-equations-steady}%
\end{subequations}
Finally, the dynamics with a conservative force that gives the same time-evolution as \eqref{moment-equations} is determined by the condition
\begin{subequations}
\begin{align}
    \bm{a}_t^* \bm{m}_t - \bm{a}^*_t  &= \bm{A}_t \bm{m}_t - \bm{a}_t , \\
    \bm{A}^*_t \bm{C}_t + \bm{C}_t \bm{A}_t^* &= \bm{A}_t \bm{C}_t + \bm{C}_t \bm{A}_t^\text{T} ,
\end{align} \label{moment-equations-optimal}%
\end{subequations}
with the constraint that $\bm{A}_t^*$ is a symmetric matrix, and where $\bm{m}_t$ and $\bm{C}_t$ are the solution of \eqref{moment-equations}.
Note that the symmetry of $\bm{A}_t^*$ is equivalent to the force being the gradient of the potential $U(\bm{x}) = \bm{x} \cdot \bm{A}_t^* \bm{x}/2 - \bm{a}^*_t \cdot \bm{x}$.
For $d = 2$, the solution of \eqref{moment-equations-optimal} can be written explicitly,
\begin{subequations}
\begin{align}
    \bm{A}_t^* &= \begin{pmatrix} K_{11} + \frac{C_{12}(A_{12} - A_{21})}{C_{11} + C_{22}} & \frac{A_{12} C_{22} + A_{21} C_{11}}{C_{11} + C_{22}} \\  \frac{A_{12} C_{22} + A_{21} C_{11}}{C_{11} + C_{22}} & A_{22} - \frac{C_{12}(A_{12} - A_{21})}{C_{11} + C_{22}}  \end{pmatrix}, \\
    \bm{a}_t^* &= \bm{a}_t + \big(\bm{A}_t^* - \bm{A}_t \big) \bm{m}_t .
\end{align}
\end{subequations}
It is easy to see that if $\bm{A}_t$ is symmetric (i.~e., the system is driven by a conservative force), then $\bm{A}_t^* = \bm{A}_t$ and $\bm{a}_t^* = \bm{a}_t$. 
The instantaneous canonical density is then given by a Gaussian with mean and covariance
\begin{subequations}
\begin{align}
    \bm{m}_t^\text{can} &= {\bm{A}_t^*}^{-1} \bm{a}_t^*, \\
    \bm{C}_t^\text{can} &= T {\bm{A}_t^*}^{-1} .
\end{align}
\end{subequations}
Using these results, we can then compute the excess, coupling and housekeeping terms in \eqref{decomposition},
\begin{subequations}
\begin{align}
    \sigma_t^\text{ex,HS} &= \mu T \bigg( \text{tr} \Big( \big( \bm{C}_t^{-1} - {\bm{C}_t^\text{st}}^{-1} \big) \bm{C}_t \big( \bm{C}_t^{-1} - {\bm{C}_t^\text{st}}^{-1} \big) \Big)  +   \Big\Vert {\bm{C}_t^\text{st}}^{-1} \big(\bm{m}_t - \bm{m}_t^\text{st} \big) \Big\Vert^2 \bigg), \\
    \sigma_t^\text{cp} &= \mu T \bigg( \text{tr} \Big( \big( {\bm{C}^\text{can}_t}^{-1} - {\bm{C}_t^\text{st}}^{-1} \big) \bm{C}_t \big( {\bm{C}^\text{can}_t}^{-1} - {\bm{C}_t^\text{st}}^{-1} \big) \Big) + \Big\Vert {\bm{C}_t^\text{st}}^{-1} \big(\bm{m}_t - \bm{m}_t^\text{st} \big) - {\bm{C}_t^\text{can}}^{-1} \big(\bm{m}_t - \bm{m}_t^\text{can} \big) \Big\Vert^2 \bigg), \\
    \sigma_t^\text{hk,MN} &= \frac{\mu}{T} \text{tr} \Big( \big(\bm{A}_t - \bm{A}_t^*\big)^\text{T} \bm{C}_t \big(\bm{A}_t - \bm{A}_t^*\big) \Big) ,
\end{align} \label{decomposition-gaussian}%
\end{subequations}
where tr denotes the trace of a matrix.
From these expression, it is immediately obvious that the excess term vanishes only when $\bm{C}_t = \bm{C}_t^\text{st}$ and $\bm{m}_t = \bm{m}_t^\text{st}$, that is, when the system is in the steady state.
On the other hand, the housekeeping term vanishes when $\bm{A}_t = \bm{A}_t^*$, that is, when the system is driven by conservative forces.
In either case, we have $p_t^\text{st}(\bm{x}) = p_t^\text{can}(\bm{x})$ and thus the coupling term vanishes as well.
The main remaining task in order to obtain explicit expressions in terms of the model parameters is solving the equations of motion \eqref{moment-equations}, which amounts to solving a set of coupled linear differential equations.
Once the solution to \eqref{moment-equations} is known, we have to solve the matrix equations \eqref{moment-equations-steady} and \eqref{moment-equations-optimal}.
We will discuss some explicit examples in Section \ref{sec-example}.
For the sake of completeness, we also provide the the expressions for the excess term of the MN decomposition and the housekeeping term of the HS decomposition,
\begin{subequations}
\begin{align}
    \sigma_t^\text{ex,MN} &= \mu T \bigg( \text{tr} \Big( \big( \bm{C}_t^{-1} - {\bm{C}_t^\text{can}}^{-1} \big) \bm{C}_t \big( \bm{C}_t^{-1} - {\bm{C}_t^\text{can}}^{-1} \big) \Big)  +   \Big\Vert {\bm{C}_t^\text{can}}^{-1} \big(\bm{m}_t - \bm{m}_t^\text{can} \big) \Big\Vert^2 \bigg) , \\
    \sigma_t^\text{hk,HS} &= \frac{\mu}{T} \bigg( \text{tr} \Big( \big( T {\bm{C}_t^\text{st}}^{-1} - \bm{K}_t \big)^\text{T} \bm{C}_t \big( T {\bm{C}_t^\text{st}}^{-1} - \bm{A}_t \big) \Big) + \Big\Vert \big( T {\bm{C}_t^\text{st}}^{-1} - \bm{A}_t \big) \Big) \big(\bm{m}_t - \bm{m}_t^\text{st} \big) \Big\Vert^2 \bigg) ,
\end{align}
\end{subequations}
as well as the expressions for the local mean velocity and its components,
\begin{subequations}
\begin{align}
    \bm{\nu}_t(\bm{x}) &= - \mu \big( \bm{A}_t \bm{x} - \bm{a}_t \big) + \mu T \bm{C}_t^{-1} \big(\bm{x} - \bm{m}_t \big), \\
    \bm{\nu}_t^\text{st}(\bm{x}) &= - \mu \big( \bm{A}_t \bm{x} - \bm{a}_t \big) + \mu T {\bm{C}_t^\text{st}}^{-1} \big(\bm{x} - \bm{m}^\text{st}_t \big), \\ 
    \bm{\nu}_t^*(\bm{x}) &= - \mu \big( \bm{A}^*_t \bm{x} - \bm{a}^*_t \big) + \mu T \bm{C}_t^{-1} \big(\bm{x} - \bm{m}_t \big) .
\end{align}
\end{subequations}

\section{Variational expressions for the HS decomposition and the coupling entropy} \label{app-variational}

\subsection{HS decomposition}
According to \eqref{orthogonality-HS} the component of the local mean velocity corresponding to the excess entropy production rate in the HS decomposition can also be written as a gradient field,
\begin{align}
\bm{\nu}_t(\bm{x}) - \bm{\nu}_t^\text{st}(\bm{x}) = - \mu T \grad \ln \bigg(\frac{p_t(\bm{x})}{p_t^\text{st}(\bm{x})} \bigg).
\end{align}
Since this is of the form \eqref{gradient-decomposition}, we immediately obtain the relation between the HS and MN housekeeping entropy from \eqref{MN-minimum},
\begin{align}
\sigma^\text{hk,HS}_t \geq \sigma_t^\text{hk,MN} .
\end{align}
However, the orthogonal complement $\bm{\nu}_t^\text{st}(\bm{x})$ is not orthogonal to arbitrary gradient fields.
Nevertheless, using the same arguments as in the derivation of \eqref{orthogonality-HS}, we can show the more general orthogonality relation
\begin{gather}
\av{\bm{v}_1, \bm{v}_2}_p = 0 \quad \text{for} \label{space-HS} \\
\bm{v}_1(\bm{x}) = \grad \phi\bigg(\frac{p_t(\bm{x})}{p_t^\text{st}(\bm{x})} \bigg) \quad \text{and} \quad \grad \cdot \big(\bm{v}_2(\bm{x}) p_t^\text{st}(\bm{x}) \big) = 0 \n  .
\end{gather}
This is similar to \eqref{space-MN}, however, $V^\text{HS}_1$ now consists of those gradient fields whose potential is a function of the ratio $p_t(\bm{x})/p_t^\text{st}(\bm{x})$, while $V^\text{HS}_2$ is the set of all vector fields which leave the instantaneous steady state $p_t^\text{st}(\bm{x})$ invariant.
Using \eqref{variational-1} and \eqref{variational-2}, we can thus obtain variational expressions for the HS excess and housekeeping entropy production rates
\begin{subequations}
\begin{align}
\sigma_t^\text{ex,HS} &= \sup_{\bm{u} \in V^\text{HS}_1} \bigg( \frac{\av{\bm{u}, \bm{\nu}_t}_p^2}{\av{\bm{v},\bm{v}}_p} \bigg) \\
&= \inf_{\bm{u} \in V^\text{HS}_2} \big( \av{\bm{\nu}_t - \bm{u}, \bm{\nu}_t - \bm{u}}_p \big) \nn
\sigma_t^\text{hk,HS} &= \inf_{\bm{u} \in V^\text{HS}_1} \big( \av{\bm{\nu}_t - \bm{u}, \bm{\nu}_t - \bm{u}}_p \big) \\
&= \sup_{\bm{u} \in V^\text{HS}_2} \bigg( \frac{\av{\bm{u}, \bm{\nu}_t}_p^2}{\av{\bm{u},\bm{u}}_p} \bigg) \n
\end{align} \label{variational-HS}%
\end{subequations}
Formally, the respective first expression resembles \eqref{variational-MN}, however, since we only optimize over functions of $p_t(\bm{x})/p_t^\text{st}(\bm{x})$, we need to know both the probability density and its instantaneous steady state value in order to evaluate the variational expressions.
From a mathematical point of view, a crucial difference between \eqref{space-MN} and \eqref{space-HS} is that, for the MN decomposition, the two subspaces $V^\text{MN}_1$ and $V^\text{MN}_2$ depend only on the probability density $p_t$ and the orthogonality condition defined by the corresponding inner product.
For the HS decomposition, by contrast, the steady state density $p_t^\text{st}(\bm{x})$ depends on the force and thus on the decomposed local mean velocity.
Thus, the definition of the orthogonal components $V^\text{HS}_1$ and $V^\text{HS}_2$ likewise depends on $\bm{\nu}_t(\bm{x})$.

We may also consider the inner product with respect to $p_t^\text{st}(\bm{x})$,
\begin{align}
\av{\bm{u},\bm{v}}_{p^\text{st}} = \frac{1}{\mu  T} \int d\bm{x} \ \bm{u}(\bm{x}) \cdot \bm{v}(\bm{x}) p_t^\text{st}(\bm{x}) \label{inner-product-steady} .
\end{align}
Since the space $V^\text{HS}_2$ consists of vector fields satisfying $\grad \cdot (\bm{v}_2(\bm{x}) p_t^\text{st}(\bm{x})) = 0$, its orthogonal space with respect to this inner product is just the space of all gradient functions $V^\text{MN}_1$.
In particular, in addition to \eqref{orthogonality-HS}, we also have
\begin{align}
\av{\bm{\nu}_t - \bm{\nu}_t^\text{st}, \bm{\nu}_t^\text{st}}_{p^\text{st}} = 0 .
\end{align}
We can thus view the HS decomposition as the MN decomposition with respect to the inner product \eqref{inner-product-steady} and the entropy production rate in the instantaneous steady state,
\begin{align}
\sigma_t^\text{st} = \av{\bm{\nu}_t^\text{st},\bm{\nu}_t^\text{st}}_{p^\text{st}},
\end{align}
also has a variational expression in terms of $\bm{\nu}_t(\bm{x})$,
\begin{align}
\sigma_t^\text{st} &= \inf_{\bm{u} \in V^\text{MN}_1} \big( \av{\bm{\nu}_t - \bm{u}, \bm{\nu}_t - \bm{u}}_{p^\text{st}} \big).
\end{align}
Together with \eqref{variational-MN-upper}, this implies the inequalities
\begin{subequations}
\begin{align}
\av{\bm{\nu}_t-\bm{\nu}_t^*,\bm{\nu}_t-\bm{\nu}_t^*}_p &\leq \av{\bm{\nu}_t^\text{st},\bm{\nu}_t^\text{st}}_p \\
\av{\bm{\nu}_t-\bm{\nu}_t^*,\bm{\nu}_t-\bm{\nu}_t^*}_{p^\text{st}} &\geq \av{\bm{\nu}_t^\text{st},\bm{\nu}_t^\text{st}}_{p^\text{st}} .
\end{align}
\end{subequations}
Thus, the lengths of the vectors $\bm{\nu}_t(\bm{x}) - \bm{\nu}_t^*(\bm{x})$ and $\bm{\nu}_t^\text{st}(\bm{x})$ satisfy opposite inequalities with respect to the two inner products.

\subsection{Coupling entropy}
In the decomposition \eqref{decomposition} into three terms, the excess part is equal to the excess part of the Hatano-Sasa decomposition, while the housekeeping part is given by the housekeeping part of the Maes-Neto{\v{c}}n{\`y} decomposition.
As a consequence, the variational expressions for the respective quantities are given by \eqref{variational-MN} and \eqref{variational-HS}.
This means that all that is left is to find the variational expression for the coupling entropy production rate $\sigma_t^\text{cp}$.
In the case of \eqref{decomposition}, we have three orthogonal components, $\bm{v}(\bm{x}) = \bm{v}_1(\bm{x}) + \bm{v}_2(\bm{x}) + \bm{v}_3(\bm{x})$.
From the preceding discussion, we can identify $V_1 = V^\text{HS}_1$ as the restricted space of gradient fields $\bm{v}_1(\bm{x}) = \grad \phi(p_t(\bm{x})/p_t^\text{st}(\bm{x}))$ (corresponding to $\bm{\nu}_t(\bm{x}) - \bm{\nu}_t^\text{st}(\bm{x})$ in the excess part), while $V_2 = V^\text{MN}_2$ is the space of vector fields that satisfy $\grad \cdot (\bm{v}_2(\bm{x}) p_t(\bm{x})) = 0$ (corresponding to $\bm{\nu}_t(\bm{x}) - \bm{\nu}_t^*(\bm{x})$ in the housekeeping part).
We then note that these two components are orthogonal due to \eqref{orthogonality-gradient}.
Then, we have to find a third component $\bm{v}_3(\bm{x})$ that is orthogonal to both $\bm{v}_1(\bm{x})$ and $\bm{v}_2(\bm{x})$.
We note that this is satisfied by
\begin{gather}
\bm{v}_3(\bm{x}) = \grad \psi(\bm{x}) = \bm{u}(\bm{x}) + \bm{w}(\bm{x}) \quad \text{with} \\
\grad \cdot \big(\bm{u}(\bm{x}) p_t(\bm{x}) \big) = 0 \quad \text{and} \quad \grad \cdot \big(\bm{w}(\bm{x}) p_t^\text{st}(\bm{x}) \big) = 0 . \n
\end{gather}
That is, the space $V_3$ consists of all gradient fields that can be written as a sum of two vector fields $\bm{u}(\bm{x})$ and $\bm{w}(\bm{x})$, which leave $p_t(\bm{x})$ and $p_t^\text{st}(\bm{x})$ invariant, respectively.
At first, it is not obvious that such vector fields exist, however, we note that one explicit example is given by
\begin{gather}
\bm{u}(\bm{x}) = \bm{\nu}_t^*(\bm{x}) - \bm{\nu}_t(\bm{x}) \nn
\bm{w}(\bm{x}) = \bm{\nu}_t^\text{st}(\bm{x}) \nn
 \Rightarrow \bm{u}(\bm{x}) + \bm{w}(\bm{x}) = - \mu T \grad \ln \bigg(\frac{p_t^\text{st}(\bm{x})}{p_t^\text{can}(\bm{x})} \bigg) .
\end{gather}
This ensures that the space $V_3$ is not empty.
Having identified the orthogonal components corresponding to \eqref{decomposition}, we have the variational expressions for the coupling entropy production rate,
\begin{align}
\sigma_t^\text{cp} &= \sup_{\bm{u} \in V_3} \bigg( \frac{\av{\bm{u},\bm{\nu}_t}_p^2}{\av{\bm{u},\bm{u}}_p} \bigg) \nn
&= \inf_{\bm{u} \in V_1 \cup V_2} \big( \av{\bm{\nu}_t - \bm{u}, \bm{\nu}_t - \bm{u}}_p \big) .
\end{align}

\section{Derivation of \eqref{xi-a} and \eqref{xi-a-reverse}} \label{app-KL-derivation}
The short-time transition probability density for the Langevin equation
\begin{align}
    \dot{\bm{x}}(t) = \mu \bm{F}_t(\bm{x}) + \bm{a}_t(\bm{x}) + \sqrt{2 \mu T} \bm{\xi}(t),
\end{align}
is given by \cite{Ris86}
\begin{align}
    p^a(\bm{x},t+dt \vert \bm{y},t) = \frac{1}{(4 \mu T dt)^\frac{d}{2}} \exp\bigg( - \frac{1}{4 \mu T dt} &\Big\Vert \bm{x} - \bm{y} - \big(\mu \bm{F}_{t + dt/2}(\bm{z}) + \bm{a}_{t+dt/2}(\bm{z}) \big) dt \Big\Vert^2 \\
    &- \frac{1}{2} \grad \cdot \big( \mu \bm{F}_{t+dt/2}(\bm{z}) + \bm{a}_{t+dt/2}(\bm{z}) \big) dt \bigg) \n ,
\end{align}
where $\bm{z} = (\bm{x}+\bm{y})/2$, and the probability density of the forward trajectory by
\begin{align}
    \mathbb{P}^a(\Gamma) = \bigg(\prod_{k=1}^M p^a(\bm{x}_{k},t_k \vert \bm{x}_{k-1},t_{k-1}) \bigg) p_0(\bm{x}_0),
\end{align}
where we defined $M = \tau/dt$ and $t_k = k dt$ and $\bm{x}_k$ denotes the position at time $t_k$.
As discussed in Section \ref{sec-fluctuation-theorem}, time-reversal involves reversing the protocol, $\bm{F}_{t}(\bm{x}) \rightarrow \bm{F}_{\tau-t}(\bm{x})$ and reversing the trajectory, $\bm{x}(t) \rightarrow \bm{x}(\tau - t)$,
\begin{align}
    \mathbb{P}^{a,\dagger}(\Gamma^\dagger) = \bigg(\prod_{k=1}^M p^a(\bm{x}_{M-k},t_{M-k} \vert \bm{x}_{M-k+1},t_{M-k+1}) \bigg) p_\tau(\bm{x}_M).
\end{align}
Note that in the time-reversed trajectory, we assume that the system starts from the final state of the original dynamics $p^{a=0}_\tau(\bm{x})$.
The corresponding path probabilities for the original dynamics are obtained by setting $\bm{a}_t(\bm{x}) = 0$.
We can then compute the logarithm of the ratio of the path probabilities,
\begin{align}
    \ln \bigg(&\frac{\mathbb{P}(\Gamma)}{\mathbb{P}^a(\Gamma)} \bigg) = \sum_{k = 1}^M \ln \bigg( \frac{p(\bm{x}_{k},t_k \vert \bm{x}_{k-1},t_{k-1})}{p^a(\bm{x}_{k},t_k \vert \bm{x}_{k-1},t_{k-1})} \bigg) \\
    &= \frac{1}{4 \mu T}  \sum_{k=1}^M \bigg( \big\Vert\bm{a}_{t + dt/2}(\bm{z}_k) \big\Vert^2 dt - 2 \big(\bm{x}_{k+1} - \bm{x}_k - \mu \bm{F}_{t+dt/2}(\bm{z}_k) \big) \cdot \bm{a}_{t+dt/2}(\bm{z}_k) + 2 \mu T \grad \cdot \bm{a}_{t+dt/2}(\bm{z}_k) dt \bigg) \n ,
\end{align}
where $\bm{z}_k = (\bm{x}_{k+1} + \bm{x}_k)/2$.
Taking the continuum limit and only keeping the leading order in $dt$, we obtain
\begin{align}
    \ln \bigg(&\frac{\mathbb{P}(\Gamma)}{\mathbb{P}^a(\Gamma)} \bigg) = \frac{1}{4 \mu T} \int_0^\tau dt \ \bigg( \big\Vert \bm{a}_t(\bm{x}(t)) \big\Vert^2 - 2 \big( \dot{\bm{x}}(t) - \mu \bm{F}_t(\bm{x}(t)) \big) \circ \bm{a}_t(\bm{x}(t)) + 2 \mu T \grad \cdot \bm{a}_t(\bm{x}(t)) \bigg) .
\end{align}
Note that here, the product between $\dot{\bm{x}}(t)$ and $\bm{a}_t(\bm{x}(t))$ has to be interpreted in the Stratonovich sense, since in the discrete-time formulation $\bm{a}_t(\bm{z}_k)$ is evaluated at the midpoint $(\bm{x}(t+dt) + \bm{x}(t))/2$.
Using the relation between the Stratonovich and Ito product,
\begin{align}
    \dot{\bm{x}}(t) \circ \bm{a}_t(\bm{x}(t)) = \dot{\bm{x}}(t) \cdot \bm{a}_t(\bm{x}(t)) + \mu T \grad \cdot \bm{a}_t(\bm{x}(t)),
\end{align}
we obtain \eqref{xi-a},
\begin{align}
    \ln \bigg(&\frac{\mathbb{P}(\Gamma)}{\mathbb{P}^a(\Gamma)} \bigg) = \frac{1}{4 \mu T} \int_0^\tau dt \ \bigg( \big\Vert \bm{a}_t(\bm{x}(t)) \big\Vert^2 - 2 \big( \dot{\bm{x}}(t) - \mu \bm{F}_t(\bm{x}(t)) \big) \cdot \bm{a}_t(\bm{x}(t))  \bigg) .
\end{align}
We remark that, in principle, the relation between the two stochastic products depends on the dynamics, that is, the concrete expression for $\bm{x}(t)$.
However, since we are only considering changes in the drift vector while keeping the diffusion coefficient constant, this relation is the same for any choice of $\bm{a}_t(\bm{x})$.
When averaging the above expression with respect to $\mathbb{P}(\Gamma)$, we can use that, along a trajectory of the original dynamics, we have
\begin{align}
    \dot{\bm{x}}(t) = \mu \bm{F}_t(\bm{x}) + \sqrt{2 \mu T} \bm{\xi}(t),
\end{align}
so that the average is given by,
\begin{align}
    \Av{\ln \bigg(\frac{\mathbb{P}(\Gamma)}{\mathbb{P}^a(\Gamma)} \bigg)} = \Av{\frac{1}{4 \mu T} \int_0^\tau dt \ \bigg( \big\Vert \bm{a}_t(\bm{x}(t)) \big\Vert^2 - 2 \sqrt{2 \mu T} \bm{\xi}(t) \cdot \bm{a}_t(\bm{x}(t))  \bigg)}.
\end{align}
Since the noise is white, the average of the second term vanishes and we have
\begin{align}
    \Av{\ln \bigg(\frac{\mathbb{P}(\Gamma)}{\mathbb{P}^a(\Gamma)} \bigg)} = \frac{1}{4 \mu T} \int_0^\tau dt \int d\bm{x} \ \big\Vert \bm{a}_t(\bm{x}) \big\Vert^2 p_t(\bm{x}),
\end{align}
which, recalling the definition of the inner product \eqref{inner-product} is precisely \eqref{xi-a-average}.
For the time-reversed path probability, we have
\begin{align}
    \ln \bigg(&\frac{\mathbb{P}(\Gamma)}{\mathbb{P}^{a,\dagger}(\Gamma)} \bigg) = \sum_{k = 1}^M \ln \bigg( \frac{p(\bm{x}_{k},t_k \vert \bm{x}_{k-1},t_{k-1})}{p^a(\bm{x}_{k-1},t_{k-1} \vert \bm{x}_{k},t_{k})} \bigg) + \ln \bigg(\frac{p_0(\bm{x}_0)}{p_\tau(\bm{x}_M)} \bigg) ,
\end{align}
where we relabelled the indices as $k \rightarrow M - k + 1$ in the product in the time-reversed path probability.
This expression can be expanded as
\begin{align}
    \ln \bigg(\frac{\mathbb{P}(\Gamma)}{\mathbb{P}^{a,\dagger}(\Gamma)} \bigg) = \frac{1}{4 \mu T}  \sum_{k = 1}^M \bigg( &\big\Vert\bm{a}_{t + dt/2}(\bm{z}_k) \big\Vert^2 dt + 2 \big( \bm{x}_{k+1} - \bm{x}_k + \mu \bm{F}_{t+dt/2}(\bm{z}_k) \big) \cdot \bm{a}_{t+dt/2}(\bm{z}_k) \\
    &\qquad + 4 \big( \bm{x}_{k+1}- \bm{x}_k \big) \cdot \mu \bm{F}_{t+dt/2}(\bm{z}_k) + 2 \mu T \grad \cdot \bm{a}_{t+dt/2}(\bm{z}_k) dt \bigg) + \ln \bigg(\frac{p_0(\bm{x}_0)}{p_\tau(\bm{x}_M)} \bigg) \n .
\end{align}
Taking the continuum limit, we obtain
\begin{align}
    \ln \bigg(\frac{\mathbb{P}(\Gamma)}{\mathbb{P}^{a,\dagger}(\Gamma)} \bigg) &= \frac{1}{4 \mu T}  \int_0^\tau dt \ \bigg( \big\Vert\bm{a}_{t}(\bm{x}(t)) \big\Vert^2 + 2 \big( \dot{\bm{x}}(t) + \mu \bm{F}_{t}(\bm{x}(t)) \big) \circ \bm{a}_{t}(\bm{x}(t)) + 4 \dot{\bm{x}}(t) \circ \mu \bm{F}_{t}(\bm{x}(t)) + 2 \mu T \grad \cdot \bm{a}_{t}(\bm{x}(t)) \bigg) \nn
    &\hspace{4cm} + \ln \bigg(\frac{p_0(\bm{x}(0))}{p_\tau(\bm{x}(\tau))} \bigg) \nn
    &= \frac{1}{4 \mu T}  \int_0^\tau dt \ \bigg( \big\Vert\bm{a}_{t}(\bm{x}(t)) \big\Vert^2 - 2 \big( \dot{\bm{x}}(t) - \bm{F}_{t}(\bm{x}(t)) \big) \cdot \bm{a}_{t}(\bm{x}(t)) + 4 \dot{\bm{x}}(t) \circ \big(\mu \bm{F}_{t}(\bm{x}(t)) + \bm{a}_t(\bm{x}(t)) \big)\bigg) \nn
    &\hspace{4cm} + \ln \bigg(\frac{p_0(\bm{x}(0))}{p_\tau(\bm{x}(\tau))} \bigg) .
\end{align}
In order to obtain \eqref{xi-a-reverse}, we use the relations
\begin{subequations}
\begin{align}
    \dot{\bm{x}}(t) \circ \mu \bm{F}_t(\bm{x}(t)) = \dot{\bm{x}}(t) \circ \big( \bm{\nu}_t(\bm{x}(t)) + \mu T \grad \ln p_t(\bm{x}(t)) \big), \\
    d_t \ln p_t(\bm{x}(t)) = \partial_t \ln p_t(\bm{x}(t)) + \grad \ln p_t(\bm{x}(t)) \circ \dot{\bm{x}}(t) ,
\end{align}
\end{subequations}
to rewrite the above as
\begin{align}
    \ln \bigg(\frac{\mathbb{P}(\Gamma)}{\mathbb{P}^{a,\dagger}(\Gamma)} \bigg) &= \frac{1}{4 \mu T}  \int_0^\tau dt \ \bigg( \big\Vert\bm{a}_{t}(\bm{x}(t)) \big\Vert^2 - 2 \big( \dot{\bm{x}}(t) - \bm{F}_{t}(\bm{x}(t)) \big) \cdot \bm{a}_{t}(\bm{x}(t)) + 4 \dot{\bm{x}}(t) \circ \big(\bm{\bm{\nu}}_{t}(\bm{x}(t)) + \bm{a}_t(\bm{x}(t)) \big) \\
    &\hspace{3cm} + 4 \mu T \big( d_t \ln p_t(\bm{x}(t)) - \partial_t \ln p_t(\bm{x}(t)) \big) \bigg) + \ln \bigg(\frac{p_0(\bm{x}(0))}{p_\tau(\bm{x}(\tau))} \bigg) . \n
\end{align}
The time-integral over the total derivative cancels the boundary term and we obtain \eqref{xi-a-reverse}
\begin{align}
    \ln \bigg(\frac{\mathbb{P}(\Gamma)}{\mathbb{P}^{a,\dagger}(\Gamma)} \bigg) &= \int_0^\tau dt \ \Bigg( \frac{1}{\mu T} \bm{\nu}_t(\bm{x}(t)) \circ \dot{\bm{x}}(t) - \partial_t \ln p_t(\bm{x}(t)) \\
    &\hspace{3cm}+ \frac{1}{4 \mu T} \bigg(  \big\Vert\bm{a}_{t}(\bm{x}(t)) \big\Vert^2 + 4 \bm{a}_t(\bm{x}(t)) \circ \dot{\bm{x}}(t) - 2 \bm{a}_t(\bm{x}(t)) \cdot \big(  \dot{\bm{x}}(t) - \bm{F}_{t}(\bm{x}(t)) \big) \bigg) \Bigg) \n .
\end{align}
The first two terms are independent of $\bm{a}_t(\bm{x})$ and correspond the stochastic entropy production \eqref{entropy-stochastic}, which is obtained by setting $\bm{a}_t(\bm{x}) = 0$.
Taking the average of this expression with respect to $\mathbb{P}(\Gamma)$, the term involving the Ito-product vanishes (see above), as does the term containing the partial derivative of the logarithm of $p_t(\bm{x})$, because we have
\begin{align}
    \int d\bm{x} \ p_t(\bm{x}) \partial_t \ln p_t(\bm{x}) = \int d\bm{x} \ \partial_t p_t(\bm{x}) = 0
\end{align}
due to conservation of probability.
Finally, when taking the average, the Stratonovich product with $\dot{\bm{x}}(t)$ becomes a product with the local mean velocity and we find
\begin{align}
    \Av{ \ln \bigg(\frac{\mathbb{P}(\Gamma)}{\mathbb{P}^{a,\dagger}(\Gamma)} \bigg)} &= \frac{1}{\mu T} \int_0^\tau dt \int d\bm{x} \ \bigg( \big(\bm{\nu}_t(\bm{x}) + \bm{a}_t(\bm{x}) \big) \cdot \bm{\nu}_t(\bm{x}) + \frac{1}{4} \big\Vert \bm{a}_t(\bm{x}) \big\Vert^2 \bigg)   p_t(\bm{x})\\
    &= \frac{1}{\mu T} \int_0^\tau dt \int d\bm{x} \ \Big\Vert \bm{\nu}_t(\bm{x}) + \frac{1}{2} \bm{a}_t(\bm{x}) \Big\Vert^2 p_t(\bm{x}), \n
\end{align}
which is \eqref{xi-a-reverse-average}.

\end{document}